\documentclass[aps,prd,onecolumn,groupedaddress,showpacs,nofootinbib,amssymb]{revtex4}
\usepackage{amssymb}
\usepackage{amsmath}
\usepackage{tensor}
\usepackage{amsfonts}
\usepackage{bm}
\usepackage{cancel}
\usepackage{comment}
\usepackage{filecontents}
\usepackage{graphicx}
\usepackage{color}
\usepackage[lofdepth,lotdepth]{subfig}
\usepackage{floatrow}
\allowdisplaybreaks[4]

\begin{document}

\title{The Phase Space of $k$-Essence $f(R)$ Gravity Theory}
\author{V.K. Oikonomou,$^{1,2,3}$\,\thanks{v.k.oikonomou1979@gmail.com}  N. Chatzarakis,$^{1}$}
\affiliation{
$^{1)}$ Department of Physics, Aristotle University of Thessaloniki, Thessaloniki 54124, Greece\\
$^{2)}$ Laboratory for Theoretical Cosmology, Tomsk State University
of Control Systems
and Radioelectronics, 634050 Tomsk, Russia (TUSUR)\\
$^{3)}$ Tomsk State Pedagogical University, 634061 Tomsk, Russia\\
}

\tolerance=5000

\begin{abstract}
Among the remaining viable theories that can successfully describe
the late-time era is the $k$-Essence theory and in this work we
study in detail the phase space of $k$-Essence $f(R)$ gravity in
vacuum. This theory can describe in a viable way the inflationary
era too, so we shall study the phase space in detail, since this
investigation may reveal general properties regarding the
inflationary attractors. By appropriately choosing the
dimensionless variables corresponding to the cosmological system,
we shall construct an autonomous dynamical system, and we find the
fixed points of the system. We focus on quasi-de Sitter
attractors, but also to radiation and matter domination
attractors, and study their stability. As we demonstrate, the
phase space is mathematically rich since it contains stable
manifold and unstable manifold. With regard to the inflationary
attractors, these exist and become asymptotically unstable, a
feature which we interpret as a strong hint that the theory has an
inherent mechanism for graceful exit from inflation. We describe
in full detail the underlying mathematical structures that control
the instability of the inflationary attractors, and also we also
address the same problem for radiation and matter domination
attractors. The whole study is performed for both canonical and
phantom scalar fields, and as we demonstrate, the canonical scalar
$k$-Essence theory is structurally more appealing in comparison to
the phantom theory, a result also demonstrated in the related
literature on $k$-Essence $f(R)$ gravity.
\end{abstract}

%PACS numbers: 04.50.Kd, 95.36.+x, 98.80.-k, 98.80.Cq

\maketitle

\section{Introduction}

The latest Planck constraints on cosmological parameters
\cite{Aghanim:2018eyx}, in conjunction with the striking
observation of gravitational waves coming from binary neutron
stars merging \cite{GBM:2017lvd}, have significantly narrowed down
the available models for the description of the late-time
acceleration era, first observed in the late 90's. Although most
modified gravity models still can describe the late-time era in a
concrete and viable way (see
\cite{Nojiri:2017ncd,Nojiri:2010wj,Nojiri:2006ri,Capozziello:2011et,Capozziello:2010zz,delaCruzDombriz:2012xy,Olmo:2011uz}
for reviews), it is still important to have alternative
descriptions that may generate a viable phenomenology. Among the
remaining viable models that can describe the dark energy era, are
the so-called $k$-Essence models
\cite{Nojiri:2019dqc,Chiba:1999ka,ArmendarizPicon:2000dh,ArmendarizPicon:1999rj,Matsumoto:2010uv,ArmendarizPicon:2000ah,Chiba:2002mw,Malquarti:2003nn,Malquarti:2003hn,Chimento:2003zf,Chimento:2003ta,Scherrer:2004au,Aguirregabiria:2004te,ArmendarizPicon:2005nz,Abramo:2005be,Rendall:2005fv,Bruneton:2006gf,dePutter:2007ny,Babichev:2007dw,Deffayet:2011gz,Kan:2018odq}
that can also describe concretely the inflationary era. It is of
crucial importance to find a model that can describe in a unified
way the dark energy era and the inflationary era. In Ref.
\cite{Nojiri:2019dqc} such a theoretical framework was given in
terms of $k$-Essence $f(R)$ gravity, and it was demonstrated that
a viable inflationary era may be generated.

However, the results of Ref. \cite{Nojiri:2019dqc} were strongly
dependent on the specific model studied, and the true structure of
the cosmological solutions must be further revealed. To this end
in this work we shall study the full phase space of the
$k$-Essence $f(R)$ gravity theory, using a simple $k$-Essence
model. In order to do so we shall construct an autonomous
dynamical system from the $k$-Essence $f(R)$ gravity theory, find
the fixed points of the cosmological system and study their
stability. We shall focus on cosmologies with physical interest,
and particularly, quasi-de Sitter fixed points, matter domination
and radiation domination fixed points. The dynamical system
approach is a very rigid and formal way to extract interesting
information about the dynamical evolution of the system and is
very frequently used in cosmology
\cite{Odintsov:2018zai,Oikonomou:2019boy,Odintsov:2018uaw,Odintsov:2018awm,Odintsov:2019ofr,Boehmer:2014vea,Bohmer:2010re,Goheer:2007wu,Leon:2014yua,Guo:2013swa,Leon:2010pu,deSouza:2007zpn,Giacomini:2017yuk,Kofinas:2014aka,Leon:2012mt,Gonzalez:2006cj,Alho:2016gzi,Biswas:2015cva,Muller:2014qja,Mirza:2014nfa,Rippl:1995bg,Ivanov:2011vy,Khurshudyan:2016qox,Boko:2016mwr,Odintsov:2017icc,Granda:2017dlx,Landim:2016gpz,Landim:2015uda,Landim:2016dxh,Bari:2018edl,Chakraborty:2018bxh,Ganiou:2018dta,Shah:2018qkh,Oikonomou:2017ppp,Odintsov:2017tbc,Dutta:2017fjw,Odintsov:2015wwp,Kleidis:2018cdx}.

In brief the results of our analysis are quite interesting since
the vacuum $f(R)$ gravity phase space is strongly stable having
stable quasi-de Sitter attractors, while the $k$-Essence $f(R)$
gravity has instabilities. Particularly, the phase space of the
$k$-Essence $f(R)$ gravity has stable inflationary attractors,
which asymptotically become unstable due to the existence of
unstable manifolds in the phase space. Eventually, the unstable
manifolds destabilize the dynamical system, and from a physical
point of view this can be viewed as graceful exit from inflation.
Similar results hold true for the radiation and matter domination
fixed points. It is notable that we performed the analysis
assuming that the scalar kinetic term corresponds to canonical
scalar fields and to phantom scalar fields, with the canonical
$k$-Essence theory showing the most physically appealing features.
In addition, we find quite intriguing substructures in the phase
space, of lower dimension in comparison to the original phase
space. These substructures control eventually the stability of the
dynamical system, these are the origin of stability. Finally, we
also examine in brief the case that no scalar kinetic term is
included, and similar results are found.

This paper is organized as follows: In section II we present in
brief the essential features of the $k$-Essence $f(R)$ gravity
model. In section III we construct the autonomous dynamical system
of the $k$-Essence $f(R)$ gravity theory, by appropriately
choosing the phase space variables, and we discuss when this
dynamical system is strictly autonomous. In section IV we
investigate in detail the phase space structure for $k$-Essence
models with canonical or phantom scalar field kinetic term, while
in section V, we study the phase space of the theory in the
absence of a scalar field kinetic term. The conclusions along with
a detailed discussion on the results, are presented in the end of
the paper.

\section{The $k$-Essence $f(R)$ Gravity Theoretical Framework}

The $k$-Essence $f(R)$ gravity theoretical framework belongs to
the general $\tilde{f}(R,\phi,X)$ theory which has the following
gravitational action,
\begin{equation} \label{hilbert}
S = \int d^{4}x \sqrt{-g} \big( \tilde{f}(R,\phi,X) + \mathcal{L}_{matter} \big) \, ,
\end{equation}
where $g^{\mu\nu}$ is the metric and $\sqrt{-g}$ its determinant.
Also $R = g^{\mu\nu} R_{\mu\nu}$ is the Ricci scalar and $X =
\dfrac{1}{2}\partial_{\mu}\phi \partial^{\mu}\phi$ is the kinetic
term of the scalar field. Finally $\mathcal{L}_{matter}$ stands
for the Lagrangian density of the matter fields. In our case we
shall assume later on that no matter fields are present, so we
will consider the vacuum case $\mathcal{L}_{matter} = 0$.

For the $k$-Essence model we shall consider, the generalized
function $\tilde{f}$ in the action has the following form,
\begin{equation} \label{actionI}
\tilde{f}(R,\phi,X) = \dfrac{1}{2 \kappa^2} f(R) + c_{1} X + G_{1} X^2 \, ;
\end{equation}
this case leads to a specific category of $k$-Essence models, to
which we shall refer to as ``Model I'' hereafter. Notice that
depending on whether $c_1=1$ or $c_1=-1$, Model I describes a
phantom scalar field or a canonical field respectively.

\subsection{Equations of Motion of the $k$-Essence $f(R)$ Gravity Theory}

Regardless of the specific form of $\tilde{f}(R,\phi,X)$ and the
coupling (or non-coupling) of the kinetic $X$ term to the Ricci
curvature, the equations of motion of the theory are derived, as
usually, by varying the gravitational action of Eq.
(\ref{hilbert}) with respect to the metric tensor, $g^{\mu\nu}$
and to the scalar field, $\phi$.  The former case yields the field
equations for the geometry of the spacetime, that is the
generalized Einstein field equations, while the latter yields the
evolution of the scalar field. We consider a flat
Friedmann-Robertson-Walker (FRW) metric of the form,
\begin{equation} \label{metric}
ds^{2} = -dt^{2} + a(t)^{2} \sum_{1}^{3} (dx^{i})^{2} \, ,
\end{equation}
where $a(t)$ is the scale factor, and thus $H(t) =
\dfrac{\dot{a}}{a}$ is the Hubble rate, and also we shall assume
that no matter fields are present, so $\mathcal{L}_{matter} = 0$.

Varying the gravitational action with respect to the metric
tensor, we obtain,
\begin{equation*}
-\dfrac{1}{2} \big( \tilde{f}(R) - R \tilde{F}(R) \big) - \dfrac{\kappa^2}{2} \tilde{f}_{X} \dot{\phi}^2 - 3 H \dot{\tilde{F}}(R) = 3 H^{2} \tilde{F}(R) \, ,
\end{equation*}
where $\tilde{F}(R) = \dfrac{\partial \tilde{f}}{\partial R}$ and
$\tilde{f}_{X} = \dfrac{\partial \tilde{f}}{\partial X}$. As a
result, since,
\begin{equation*}
\ddot{\tilde{F}}(R) = -\dfrac{1}{2}\tilde{f}(R) + \tilde{F}(R) \Big( \dfrac{R}{2} - H^{2} \Big) - 2 H \dot{\tilde{F}}(R) \, ,
\end{equation*}
the field equations for the $f(R)-X$ theory become
\begin{equation} \label{fieldeq}
\ddot{\tilde{F}}(R) - H \dot{\tilde{F}}(R) - \dfrac{\kappa^2}{2} \tilde{f}_{X} \dot{\phi}^2 = 0 \, .
\end{equation}
Varying the gravitational action with respect to the scalar field,
we obtain,
\begin{equation*}
\dfrac{1}{a^3} \big( a^3 \tilde{f}_{X} \dot{\phi} \big)^{\cdot} + \tilde{f}_{\phi} = 0 \, ,
\end{equation*}
where $\tilde{f}_{\phi} = \dfrac{\partial \tilde{f}}{\partial
\phi}$. Hence, the equation of motion for the scalar field is,
\begin{equation} \label{fieldmotion}
\tilde{f}_{X} \big( \ddot{\phi} + 3 H \dot{\phi} \big) + \dot{\tilde{f}}_{X}\dot{\phi} + \tilde{f}_{\phi} = 0 \, .
\end{equation}

We can now specify the equations of motion for Model I given in
Eq. (\ref{actionI}). Since $\tilde{f}_{I}(R,\phi,X) = \dfrac{1}{2
\kappa^2} f(R) + c_{1} X + G_{1} X^2$, we have,
\begin{equation*}
\tilde{F}_{I}(R) = \dfrac{\partial \tilde{f}_{I}}{\partial R} =
\dfrac{1}{2 \kappa^2} F(R) \;\; \, , \;\;\; \tilde{f}_{I,\phi} =
\dfrac{\partial \tilde{f}_{I}}{\partial \phi} = 0 \;\; \text{and}
\;\;\; \tilde{f}_{I,X} = \dfrac{\partial \tilde{f}_{I}}{\partial
X} = c_{1} + 2 G_{1} X \, .
\end{equation*}
where $c_{1}$ and $G_{1}$ are constants and can be viewed as free
parameters for the models. Concerning $c_1$, its sign can indicate
the type of scalar field cosmologies used, with $c_{1} = -1$
describing a canonical scalar field, while $c_{1} = 1$ describing
phantom scalar fields, and $c_{1} = 0$ denoting the absence of a
kinetic term, which is physically unmotivated, though we will
examine this case as well. Consequently, the field equation
becomes,
\begin{equation} \label{fieldeqI}
\ddot{F}(R) - H \dot{F}(R) - \kappa^4 \dot{\phi}^2 \big( c_{1} + 2 G_{1} X \big) = 0 \, ,
\end{equation}
and the evolution of the scalar field,
\begin{equation} \label{fieldmotionI}
\big( c_{1} +2 G_{1} X \big) \big( \ddot{\phi} + 3 H \dot{\phi} ) + 2 G_{1} \dot{X}\dot{\phi} = 0 \, .
\end{equation}
Having the equations of motion at hand, we can introduce several
dimensionless dynamical variables, and we shall construct an
autonomous dynamical system, the phase space of which we shall
extensively study.

\section{Setting up the Dynamical Model}

In order to examine the cosmological implications and behavior of
Model I (\ref{actionI}), we shall investigate the mathematical
structure of its phase space. In order to do so, we need to
introduce appropriate dimensionless phase space variables which
will constitute an autonomous dynamical system. Taking into
account that in a flat FRW spacetime, we have,
\begin{align}
R &= 6 \dot{H} + 12 H^{2} \;\;\; \text{and} \;\;\; \dot{R} = 24\dot{H} H + 6 \ddot{H} \\
X &= -\dfrac{1}{2} \dot{\phi}^2 \;\;\; \text{and} \;\;\; \dot{X} = -\dot{\phi} \ddot{\phi} \; \, ,
\end{align}
we define the following five dimensionless phase space variables,
\begin{equation} \label{statevar}
x_{1} = - \dfrac{\dot{F}}{H F} \;\; \, , \;\;\; x_{2} = - \dfrac{f}{6 H^2 F} \;\; \, , \;\;\; x_{3} = \dfrac{R}{6 H^2} \;\; \, , \;\;\; x_{4} = \kappa^2 \dot{\phi} \;\;\; \text{and} \;\;\; x_{5} = \dfrac{1}{\kappa^2 H^2 F} \, .
\end{equation}
The first three of these variables are typical and have been
defined as such in many similar $f(R)$ gravity phase space studies
\cite{Odintsov:2017tbc}, however the variables $x_{4}$ and
$x_{5}$, are needed only in the $k$-Essence $f(R)$ gravity case.
Their evolution will be studied by using the $e$-foldings number,
$N$, defined as follows,
\begin{equation} \label{e-folds}
N = \int_{t_{in}}^{t_{fin}} H(t) dt \, ,
\end{equation}
where $t_{in}$ and $t_{fin}$ the initial and final time instances.
The derivatives in respect to the $e$-foldings number are derived
from the derivatives with respect to time, by using,
\begin{equation*}
\dfrac{d}{dN} = \dfrac{1}{H} \dfrac{d}{dt}\, , \;\;  \;\;\;
\dfrac{d^2}{dN^2} = \dfrac{1}{H} \Big( \dfrac{d^2}{dt^2} -
\dfrac{\dot{H}}{H} \dfrac{d}{dt} \Big) \, .
\end{equation*}
The equations governing the evolution of the five variables with
respect to the $e$-foldings number are given from the equations of
motion, expressed in terms of these variables.

Specifically, the evolution of $x_{1}$ with respect to the
$e$-foldings number is given as,
\begin{equation}
\dfrac{d x_{1}}{dN} = \dfrac{1}{H} \dot{x}_{1} = -\dfrac{1}{H^2} \Bigg( \dfrac{\ddot{F}}{F} - \dfrac{\dot{H}}{H} \dfrac{\dot{F}}{F} - \Big( \dfrac{\dot{F}}{F} \Big)^2 \Bigg) \, ,
\end{equation}
where, from the field equation (Eq. (\ref{fieldeqI})), we derived,
\begin{equation*}
-\dfrac{\ddot{F}}{H^{2} F} = -\dfrac{\dot{F}}{H F} - \dfrac{\kappa^{2} \dot{\phi}^{2}}{2 H^{2} F} \big( c_{1} - G_{1} \dot{\phi}^{2} \big) = x_{1} -\dfrac{c_{1} - f_{D} x_{4}^{2}}{2} x_{4}^{2} x_{5} \, ,
\end{equation*}
where $f_{D} = \dfrac{G_{1}}{\kappa^{4}}$ so that it becomes
dimensionless, and also we used the definition of the variables,
\begin{equation*}
\dfrac{\dot{H}}{H^{3}} \dfrac{\dot{F}}{F} = - x_{1} (x_{3} - 2)\,
, \;\;\; \;\;\; \Big( \dfrac{\dot{F}}{H F} \Big)^{2} = - x_{1}^{2}
\, .
\end{equation*}
Consequently, the differential equation describing the evolution
of $x_1$ is equal to,
\begin{equation} \label{firstODE}
\dfrac{d x_{1}}{dN} = - 4 + 3 x_{1} + x_{1}^2 - x_{1}x_{3} + 2 x_{3} - \dfrac{c_{1} - f_{D} x_{4}^2}{2} x_{4}^2 x_{5} \, .
\end{equation}

The evolution of $x_{2}$ with respect to the $e$-foldings number
is given as,
\begin{equation}
\dfrac{d x_{2}}{dN} = \dfrac{1}{H} \dot{x}_{2} = \dfrac{f}{6 H^3 F} \Big( -\dfrac{\dot{f}}{f} + \dfrac{\dot{H}}{H} + \dfrac{\dot{H}}{H} \Big) \, ,
\end{equation}
where, we used the definition of the phase space variables,
\begin{align*}
-\dfrac{\dot{f}}{6 H^3 F} &= -\dfrac{\dot{R}}{6 H^3} = - 4(x_{3} - 2) - m \;\; \, , \\
\frac{f}{6 H^2 F} \dfrac{\dot{H}}{H^2} &= -2 x_{2} (x_{3} - 2)\, ,
\;\;\;  \;\;\; \dfrac{f}{6 H^2 F} \dfrac{\dot{F}}{H F} = - x_{1}
x_{2} \, ,
\end{align*}
where $m = -\dfrac{\ddot{H}}{H^3}$ is a dynamical variable of
crucial importance. In our study, the dynamical system we shall
derive will be autonomous only in the case that the variable $m$
is constant. Thus, the resulting differential equation describing
the evolution of the variable $x_2$ is the following,
\begin{equation} \label{secondODE}
\dfrac{d x_{2}}{dN} = 8 - m + 4 x_{2} - 4 x_{3} + x_{1}x_{2} - 2 x_{2}x_{3} \, .
\end{equation}

Accordingly, the evolution of $x_{3}$ with respect to the
$e$-foldings number is given as follows,
\begin{equation}
\dfrac{d x_{3}}{dN} = \dfrac{1}{H} \dot{x}_{3} = \dfrac{R}{6 H^3}
\Big( \dfrac{\dot{R}}{R} - \dfrac{\dot{H}}{H} \Big) \, ,
\end{equation}
From the definition of the phase space variables, we obtain,
\begin{equation*}
\dfrac{\dot{R}}{6 H^3} = 4 \dfrac{\dot{H}}{H^2} + \dfrac{\ddot{H}}{H^3} = 4 (x_{3} - 2) - m \;\;\; \text{and} \;\;\; \dfrac{R}{6 H^2} \dfrac{\dot{H}}{H^2} = x_{3} (x_{3} - 2) \, .
\end{equation*}
Consequently, the third differential equation is,
\begin{equation} \label{thirdODE}
\dfrac{d x_{3}}{dN} = - 8 - m + 8 x_{3} - 2 x_{3}^2 \, .
\end{equation}
We should notice that this differential equation is independent
from the rest, since the evolution of $x_{3}$ depends only on the
variable itself and the parameter $m$. We will show later that
this differential equation can be solved analytically for constant
$m$.

The evolution of $x_{4}$ depends strongly on the value of the
parameter $c_{1}$, in such a way that the differential equation
governing the evolution has a completely different form when
$c_{1} = 0$. The evolution is given by,
\begin{equation}
\dfrac{d x_{4}}{dN} = \dfrac{1}{H} \dot{x_{4}} = \dfrac{\kappa^2 \ddot{\phi}}{H} \, .
\end{equation}
The second temporal derivative of the scalar field $\phi$ is
derived from the equation of motion for the respective field,
namely Eq. (\ref{fieldmotionI}), whose form changes drastically
with $c_{1} = 0$. To demonstrate this, we substitute $X= -
\dfrac{1}{2} \dot{\phi}^2$ and $\dot{X} = - \dot{\phi}\ddot{\phi}$
in Eq. (\ref{fieldmotionI}) and solve it with respect to
$\ddot{\phi}$, obtaining,
\begin{equation} \label{field2der}
\ddot{\phi} = - 3 H \dot{\phi} \dfrac{c_{1} - f_{D} \kappa^4
\dot{\phi}^2}{c_{1} - 3 f_{D} \kappa^4 \dot{\phi}^2} \, .
\end{equation}
When $c_{1} = 0$, this expression is simplified to
\begin{equation} \label{specialfield2der}
\ddot{\phi} = - H \dot{\phi} \, .
\end{equation}
In effect, we should consider two distinct forms of Model I,
hereafter called ``Model I$\alpha$'' and ``Model I$\beta$'', with
the first describing the case $c_{1} \neq 0$ and the second
describing the case $c_{1} = 0$. The specific forms of the fourth
differential equation are given below:

\begin{itemize}

\item[1] For $c_{1} \neq 0$, the differential equation describing
the evolution of the phase space variable $x_4$ becomes,
\begin{equation} \label{fourthODEa}
\dfrac{d x_{4}}{dN} = \dfrac{3}{3 f_{D} x_{4}^2 - c_{1}} \big( x_{4} - x_{4}^3 \big) \, .
\end{equation}
This differential equation resembles a non-linear oscillation,
multiplied with a damping term. Similar to Eq. (\ref{thirdODE}),
the differential equation (\ref{fourthODEa}) is independent from
all the other phase space variables and can be analytically
integrated, as we will show shortly.

\item[2] For $c_{1} = 0$, the differential equation describing the
evolution of the phase space variable $x_4$ becomes,
\begin{equation} \label{fourthODEb}
\dfrac{d x_{4}}{dN} = - x_{4} \, .
\end{equation}
This differential equation leads to a simple exponential
evolution. Obviously, it is also independent and analytically
integrated.

\end{itemize}
Finally, the evolution of the variable $x_{5}$ is given as,
\begin{equation}
\dfrac{d x_{5}}{dN} = \dfrac{1}{H} \dot{x}_{5} = - \dfrac{1}{\kappa^2 H^3 H} \Big( \dfrac{\dot{F}}{F} - 2 \dfrac{\dot{H}}{H} \Big) \, .
\end{equation}
From the definition of the phase space variables, we may transform
it to
\begin{equation} \label{fifthODE}
\dfrac{d x_{5}}{dN} = \big( 4 - x_{1} + 2x_{3} \big) x_{5} \, .
\end{equation}
In conclusion, the dynamical system of the $k$-Essence $f(R)$
gravity model (\ref{actionI}) corresponding to $c_1=-1$ and $c_1=1$ is the
following,
\begin{align}\label{dynsystem1}
& \dfrac{d x_{1}}{dN} = - 4 + 3 x_{1} + x_{1}^2 - x_{1}x_{3} + 2
x_{3} - \dfrac{c_{1} - f_{D} x_{4}^2}{2} x_{4}^2 x_{5} \, , \\
\notag & \dfrac{d x_{2}}{dN} = 8 - m + 4 x_{2} - 4 x_{3} +
x_{1}x_{2} - 2 x_{2}x_{3} \, , \\
\notag & \dfrac{d x_{3}}{dN} = - 8 - m + 8 x_{3} - 2 x_{3}^2 \,
,\\ \notag & \dfrac{d x_{4}}{dN} = \dfrac{3}{3 f_{D} x_{4}^2 -
c_{1}} \big( x_{4} - x_{4}^3 \big) \, , \\ \notag & \dfrac{d
x_{5}}{dN} = \dfrac{1}{H} \dot{x}_{5} = \big( 4 - x_{1} + 2x_{3}
\big) x_{5} \, .
\end{align}
while the one corresponding to $c_1=0$ is equal to,
\begin{align}\label{dynsystem2}
& \dfrac{d x_{1}}{dN} = - 4 + 3 x_{1} + x_{1}^2 - x_{1}x_{3} + 2
x_{3} - \dfrac{c_{1} - f_{D} x_{4}^2}{2} x_{4}^2 x_{5} \, , \\
\notag & \dfrac{d x_{2}}{dN} = 8 - m + 4 x_{2} - 4 x_{3} +
x_{1}x_{2} - 2 x_{2}x_{3} \, , \\
\notag & \dfrac{d x_{3}}{dN} = - 8 - m + 8 x_{3} - 2 x_{3}^2 \, ,
\\ \notag & \dfrac{d x_{4}}{dN} = - x_{4} \, , \\ \notag & \dfrac{d
x_{5}}{dN} = \dfrac{1}{H} \dot{x}_{5} = \big( 4 - x_{1} + 2x_{3}
\big) x_{5} \, .
\end{align}
In the following sections we shall extensively study the above
dynamical systems in detail.

\subsection{Friedmann Constraint and the Effective Equation of State}

Considering all ingredients of the Universe as homogeneous ideal
fluids, we may write down an effective equation of state (EoS) as
follows,
\begin{equation*}
w_{eff} = \dfrac{P}{\rho} \, ,
\end{equation*}
where $\rho$ is the energy density of the matter fields and $P$ is
the corresponding isotropic pressure. The effective barotrobic
index, $w_{eff}$, is equal to,
\begin{equation}
w_{eff} = - 1 - \dfrac{2}{3} \dfrac{\dot{H}}{H^2} \, .
\end{equation}
Given that the Ricci scalar in a FRW space time is $R = 6 \dot{H}
+ 12 H^2$, and, from the definitions of the phase space variables,
$x_{3} = \dfrac{R}{6 H^2} = \dfrac{\dot{H}}{H^2} + 2$, we have
\begin{equation} \label{effeqstate}
w_{eff} = - \dfrac{1}{3} (2 x_{3} - 1) \, .
\end{equation}
The effective equation of state must be satisfied by all the fixed
points of the dynamical systems (\ref{dynsystem1}) and
(\ref{dynsystem2}), if these fixed points are physical.

By looking Eq. (\ref{effeqstate}), it is apparent that $x_{3}$
determines the value of the EoS parameter $w_{eff}$ in the
following way,
\begin{itemize}
\item[1] If the Universe is in a de Sitter expansion phase, so
that $w_{eff} = -1$, then $x_{3} = 2$. \item[2] If the Universe is
dominated by effective curvature, so that $w_{eff} =
-\dfrac{1}{3}$, then $x_{3} = 1$. \item[3] If the Universe is
dominated by a pressure-free non-relativistic fluid (dust) so that
$w_{eff} = 0$, which corresponds to the matter-dominated era, then
$x_{3} = \dfrac{1}{2}$. \item[4] If the Universe is dominated by a
relativistic fluids that $w_{eff} = \dfrac{1}{3}$, resulting to
the radiation-dominated era, than $x_{3} = 0$. \item[5] If the
Universe is dominated by stiff matter so that $w_{eff} = 1$, then
$x_{3} = -1$.
\end{itemize}

Another important relation that needs to be fulfilled by the model
is the Friedman constraint, derived from the Friedman equation.
Writing down the Friedman equation as,
\begin{equation}
- \dfrac{\dot{F}}{H F} - \dfrac{f}{6 H^2 F} + \dfrac{R}{6 H ^2} - \dfrac{\kappa^2}{2} \dfrac{f_{X} \dot{\phi}}{3 H^2 F} = 1 \, ,
\end{equation}
where the first three terms correspond to the curvature and the
fourth to the scalar field, and by using the definition of the
phase space variables, we obtain,
\begin{equation}
x_{1} + x_{2} + x_{3} - \dfrac{f_{X}}{6} x_{4}^2 x_{5} = 1 \, .
\end{equation}
Apparently, the constraint depends on the form of $\tilde{f}_{I}$.
Since $f_{X} = c_{1} + 2 G_{1} X = c_{1} - f_{D} x_{4}^2$ and
$c_{1}$ may come with two distinct versions of Model I, we have
the following two cases:

\begin{itemize}

\item[1] When $c_{1} \neq 0$, so we refer to Model I$\alpha$, then
$f_{X} = c_{1} - f_{D} x_{4}^2$. The Friedman constraint for Model
I$\alpha$ is the following,
\begin{equation} \label{FriedconstrainIa}
x_{1} + x_{2} + x_{3} - \dfrac{c_{1} - f_{D} x_{4}^2}{6} x_{4}^2 x_{5} = 1 \, .
\end{equation}

\item[2] In the special case of $c_{1} = 0$, when we refer to
Model I$\beta$, then $f_{X} = - f_{D} x_{4}^2$. Hence, the
Friedman constraint for Model I$\beta$ becomes,
\begin{equation} \label{FriedconstrainIb}
x_{1} + x_{2} + x_{3} + \dfrac{f_{D}}{6} x_{4}^4 x_{5} = 1 \, .
\end{equation}

\end{itemize}
Having the above at hand, in the next sections we proceed to the
analysis of the phase space structure for the $k$-Essence $f(R)$
gravity models we discussed in the previous sections.

As for parameter $m$, given the fact that it is merely the ratio
of the second derivative over the cube of the Hubble rate, it is
subject to the specific nature of the spacetime, that is of the
specific nature of its matter content. Consequently, the values of
$m$ are also very specific, if this is considered to be constant,
which is our case. If we consider the three major phases of cosmic
evolution, namely the quasi-de Sitter expansion, the
matter-dominated era and the radiation-dominated era, we know that
the Hubble rate has specific functional forms. Specifically, in
the case of  quasi-de Sitter expansion, and $H(t)=H_0-H_it$, thus
$m = 0$. In the case of matter domination, $H(t) = \dfrac{2}{3
t}$, and thus $m = - \dfrac{9}{2}$ and in the case of radiation
domination, $H(t) = \dfrac{1}{2t}$, and thus $m = -8$. These
values of $m$ are the values with major cosmological interest,
hence the only values to be used hereafter.
%%%%%%%%%%%%%%%%%%%%%%%%%%%%%%%%%%%%%%%%%%%%%%%%%%%%%%%%%%%%%%%%%%%%%%%%%%%%%%%%%%%%%%%%%%%%%%%%%%%%%%%%%%%%%%thursday 02-05

\subsection{Integrability of the Differential Equations for $x_3$ and $x_4$}

As we noted earlier, two of the differential equations composing
the dynamical systems (\ref{dynsystem1}) and (\ref{dynsystem2}),
namely Eqs. (\ref{thirdODE}) and (\ref{fourthODEa}) for Model
I$\alpha$ and Eqs. (\ref{thirdODE}) and (\ref{fourthODEb}) for
Model I$\beta$, are independent from the other three, meaning that
they do not contain any other phase space variables. Consequently,
these first-order differential equations can be solved
independently  from the other, and it proves that these can be
solved analytically. The independence of these equations, and by
extent of the behavior of these two phase space variables, as well
as the analytical solutions derived from them, can in fact explain
the symmetries we observe later in the corresponding values of
$x_{3}$ and $x_{4}$ in the equilibria, for all the cases we shall
study (for any value of $m$ and $f_{D}$). They can also explain
the stability properties these equilibrium values have, and the
corresponding behavior of these variables independently of the
others.

\begin{figure}[h!]
\centering
\includegraphics[width=18pc]{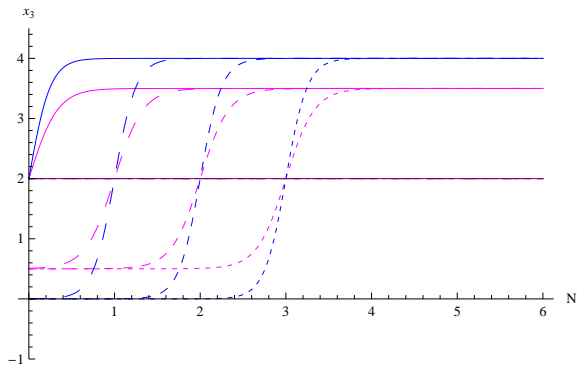}
\caption{Analytical solutions derived for the differential
equation (\ref{thirdODE}) for different initial values (solid
curves correspond to $N_{0}= 0$, while thinner dashing corresponds
to greater $N_{0}>0$), for $m=0$ (purple curves), $m=-\frac{9}{2}$
(magenta curves) and $m=-8$ (blue curves). The fast convergence to
some equilibrium value is easily observable within 5 to 6
$e$-foldings.} \label{fig:analytical_x3}
\end{figure}

Beginning with Eq. (\ref{thirdODE}), the analytical solution is
\begin{equation} \label{x3_analytical}
x_{3}(N) = 2 - \dfrac{ \sqrt{-2m} }{2} \tan \big( \sqrt{-2m} (N - N_{0}) \big) \, ,
\end{equation}
where $N_{0}$ the e-foldings number corresponding to the initial
time, \textit{i.e.} the integration constant of Eq.
(\ref{e-folds}) -usually chosen to be $N_{0} = 0$ for simplicity,
in the case of inflationary evolutions, since $t_{in} = t_{Pl}
\simeq 5.39 \; 10^{-44}$ the initial moment for inflation.
Assuming that $m=$constant, the differential equation
(\ref{thirdODE}) has no equilibrium points for $m > 0$, and has
one stable equilibrium point for $m = 0$, $x_{3}^{*} = 2$, and two
equilibrium points for $m < 0$, namely, $x_{3}^{*} = 2 \pm \dfrac{
\sqrt{-2m} }{2}$. Of the two, $x_{3}^{*} < 2$ is the unstable and
$x_{3}^{*} > 2$ is the stable equilibrium point. As a result, the
analytical solutions for $m = 0$ are $x_{3} = 2$ for any $N$, and
for $m < 0$, the solutions converge rather fast to $x_{3}^{*} = 2
+ \dfrac{ \sqrt{-2m} }{2}$, with the rate of convergence depending
on the initial conditions, typically on $N_{0}$. In Fig.
\ref{fig:analytical_x3} we have plotted the behavior of the
variable $x_3$ for various values of the parameter $m$, namely for
$m=0$ (quasi-de Sitter cosmologies), $m=-9/2$ (matter domination
cosmology) and $m=-8$ (radiation domination cosmology).

\begin{figure}[h!]
\centering
\includegraphics[width=18pc]{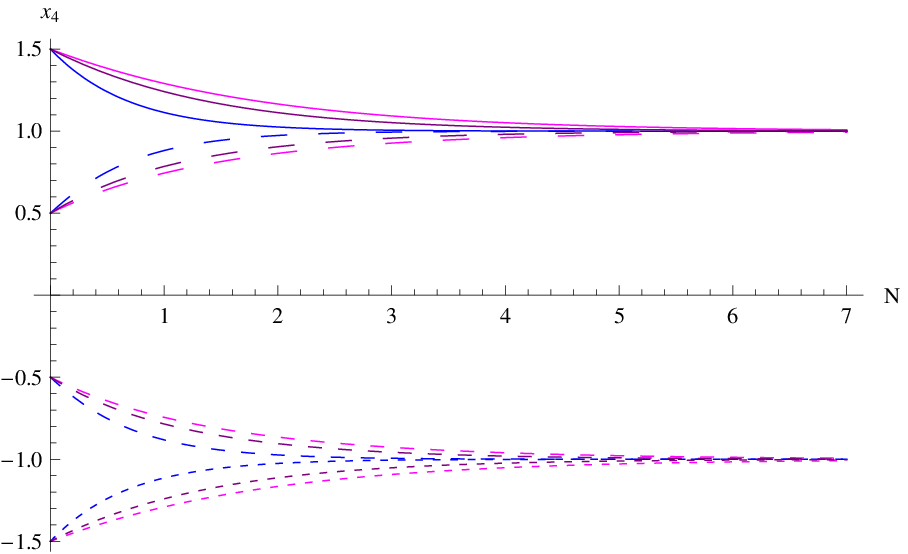}
\includegraphics[width=18pc]{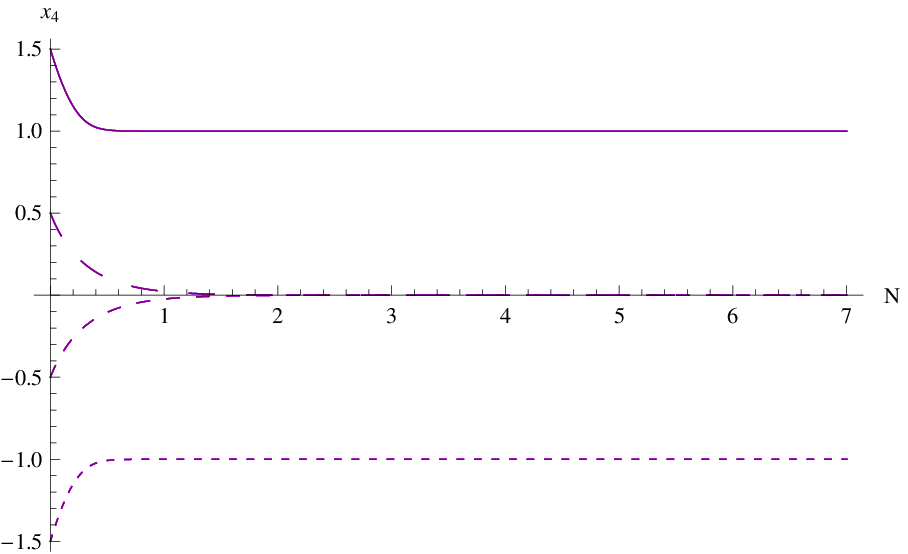}
\caption{Analytical solutions derived for the differential
equation (\ref{fourthODEa}) for different initial values (solid
curves correspond to $N_{0} = 0$, while thinner dashing
corresponds to greater $N_{0} >0$), for $m=0$ (purple curves),
$m=-\dfrac{9}{2}$ (magenta curves) and $m=-8$ (blue curves), and
for both cases ($c_{1} = -1$ on the left, and $c_{1} = 1$ on the
right). The fast convergence is easily observable within $5$ to
$6$ e-foldings.} \label{fig:analytical_x4}
\end{figure}

As for Eq. (\ref{fourthODEb}), the solution is,
\begin{equation}
x_{4} = x_{4}(0) e^{-N} \, ,
\end{equation}
where $x_{4}(0)$ determined by the initial conditions. This
solution converges asymptotically, though rapidly to $x_{4}^{*} =
0$, which is proved to be the equilibrium value for $c_{1} = 0$.
It is remarkable that this convergence does not depend on any
other parameter.

Finally, the analytical solution of the Eq. (\ref{fourthODEa}) is
generally derived by means of inverse functions, so it is not
possible to present it in closed form. It is however easy to plot
it and extract the general behavior, and in Fig.
\ref{fig:analytical_x4} we present the behavior of $x_4$ for
various $m$ and $c_1$. What is interesting about it, is that Eq.
(\ref{fourthODEa}) has three equilibrium points, $x_{4}^{*} = 0$,
$x_{4}^{*} = -1$ and $x_{4}^{*} = -1$, with the first being
unstable, and the other two stable for $c_{1} = -1$, while all are
stable for $c_{1} = 1$. Consequently, for $c_{1} = -1$ (canonical
scalar cosmologies), the solutions with positive initial values
converge rather fast to $x_{4}^{*} = 1$ and those with negative
initial values converge equally fast to $x_{4}^{*} = -1$. For
$c_{1} = -1$ (cosmologies with canonical scalar fields), the
solutions with initial values larger than unity converge to
$x_{4}^{*} = 1$, those with initial values smaller than $-1$
converge to $x_{4}^{*} = -1$ and, finally, those with initial
values in the interval $[-1,1]$ converge to $x_{4}^{*} = 0$. The
rate of the convergence in any of these cases, depends on the
choice of $f_{D}$. In Fig. \ref{fig:analytical_x4} we present the
behavior of the variable $x_4$ for various values of the parameter
$m$ and $c_1$.

\subsection{The three free parameters}

One small comment is needed for the free parameters of the two
models. Apart from $c_{1}$ that is inherent from the original
theory, two more are included, defined as,
\begin{equation}
m = -\dfrac{\ddot{H}}{H^3}\, , \;\;\;  \;\;\; f_{D} =
\dfrac{G_{1}}{\kappa^4} \, .
\end{equation}
Generally, the existence of one free parameter means that the
corresponding dynamical system might pass through a number of
bifurcations, accordingly to its dependence on this free
parameter. The existence of more than one could make the situation
far more complicated, with many more bifurcations occurring as
different values can be assigned to all the free parameters. This
essentially means that the structure of the phase space, and
consequently the behavior of the system, may change. The existence
of fixed points is questioned, their stability might be altered,
attractors can appear and disappear, even chaotic behavior may
arise.

Hopefully, in our case things are quite simpler, since our three
free parameters are not so arbitrarily chosen actually. Due to the
role they play in the model under study, they can be given very
specific values. As a result, the parameter space is contained and
certain aspects of the bifurcation analysis are similar. More
specifically, $c_{1}$ can take just two values for Model
I$\alpha$, that are $-1$ and $1$, and only one for Model I$\beta$,
that is $0$; the reasons for this have been already explained
earlier.

\begin{figure}[h!] % "[t!]" placement specifier just for this example
\includegraphics[width=18pc]{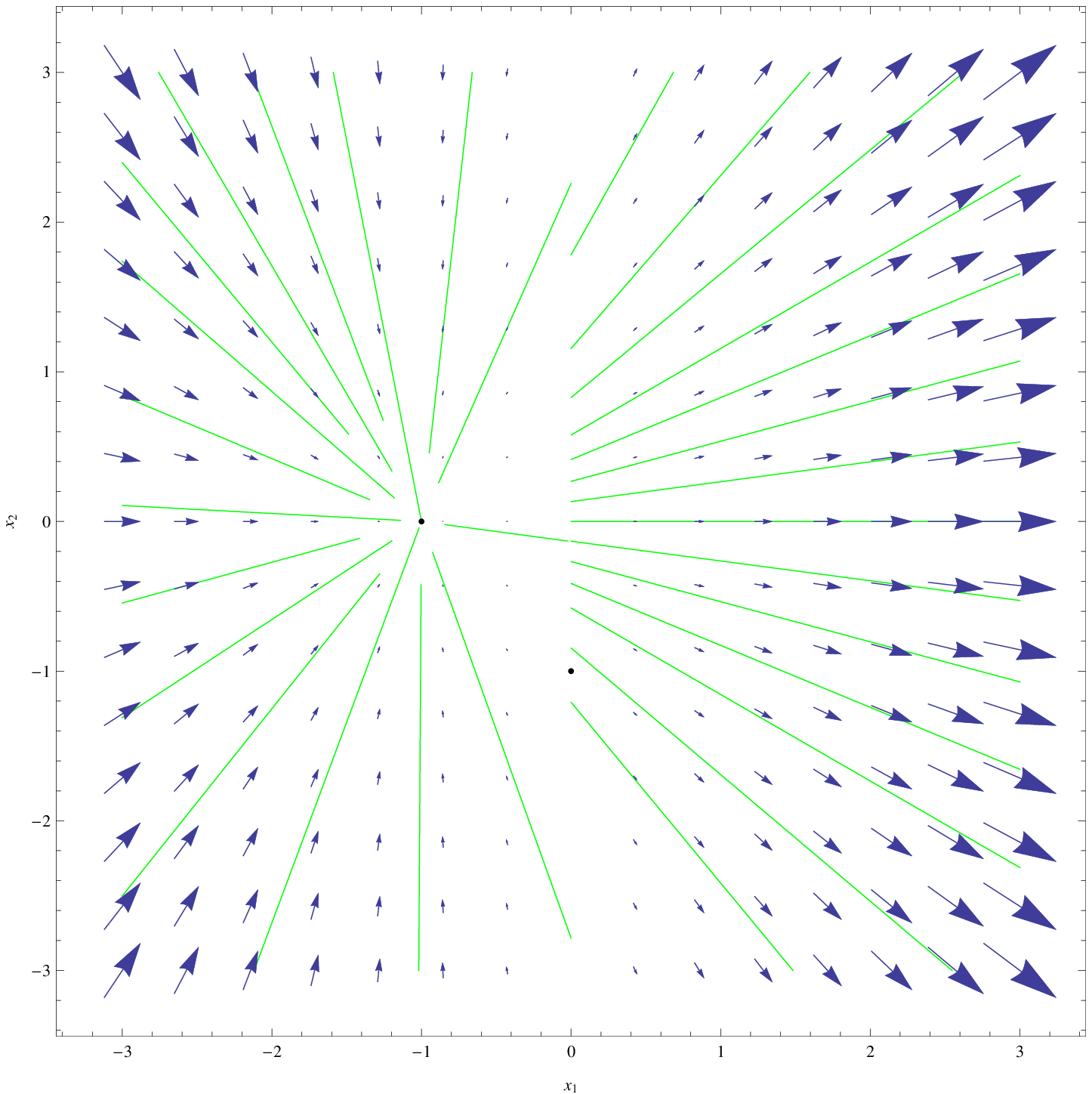}
\includegraphics[width=18pc]{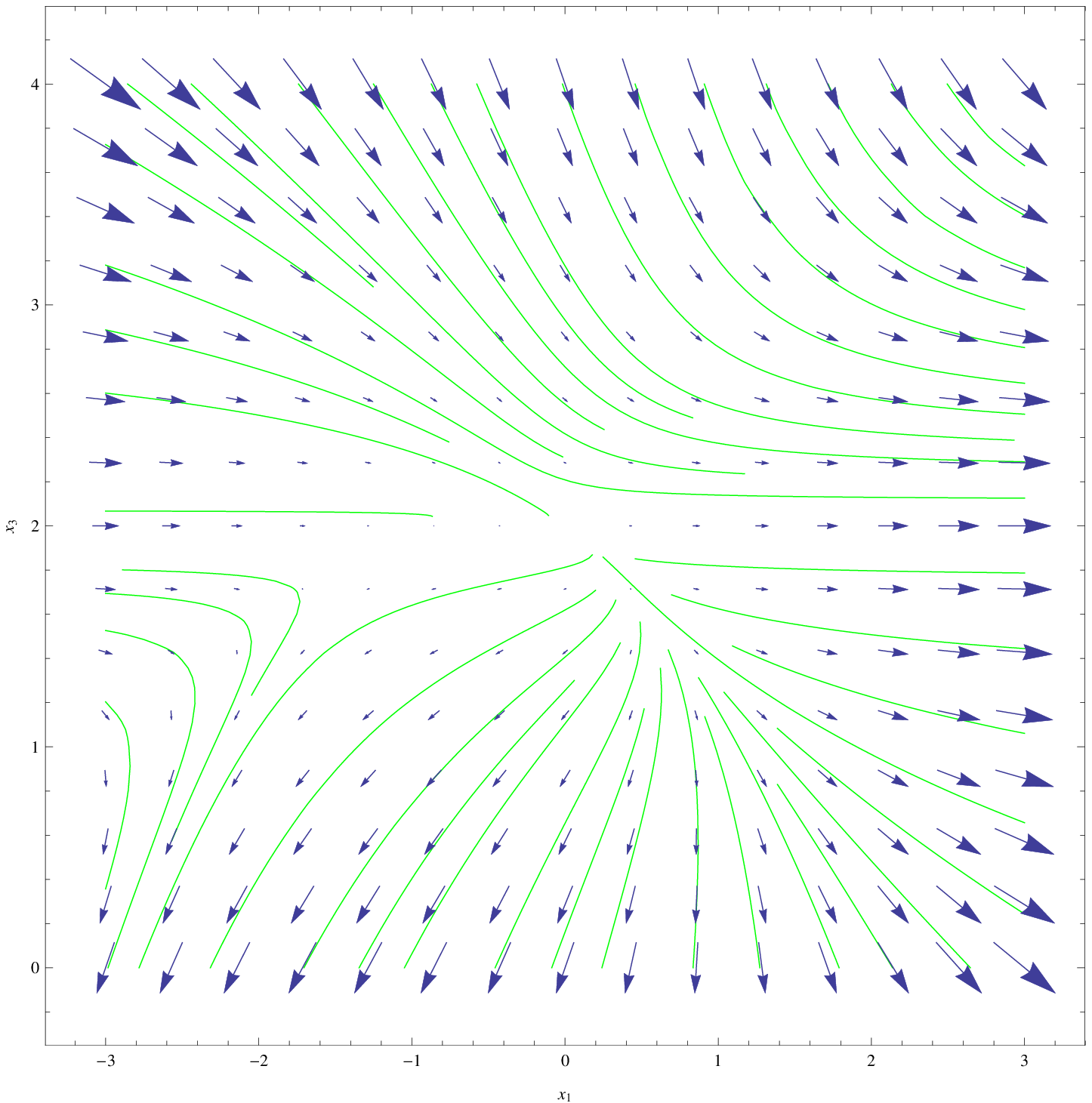}
\includegraphics[width=18pc]{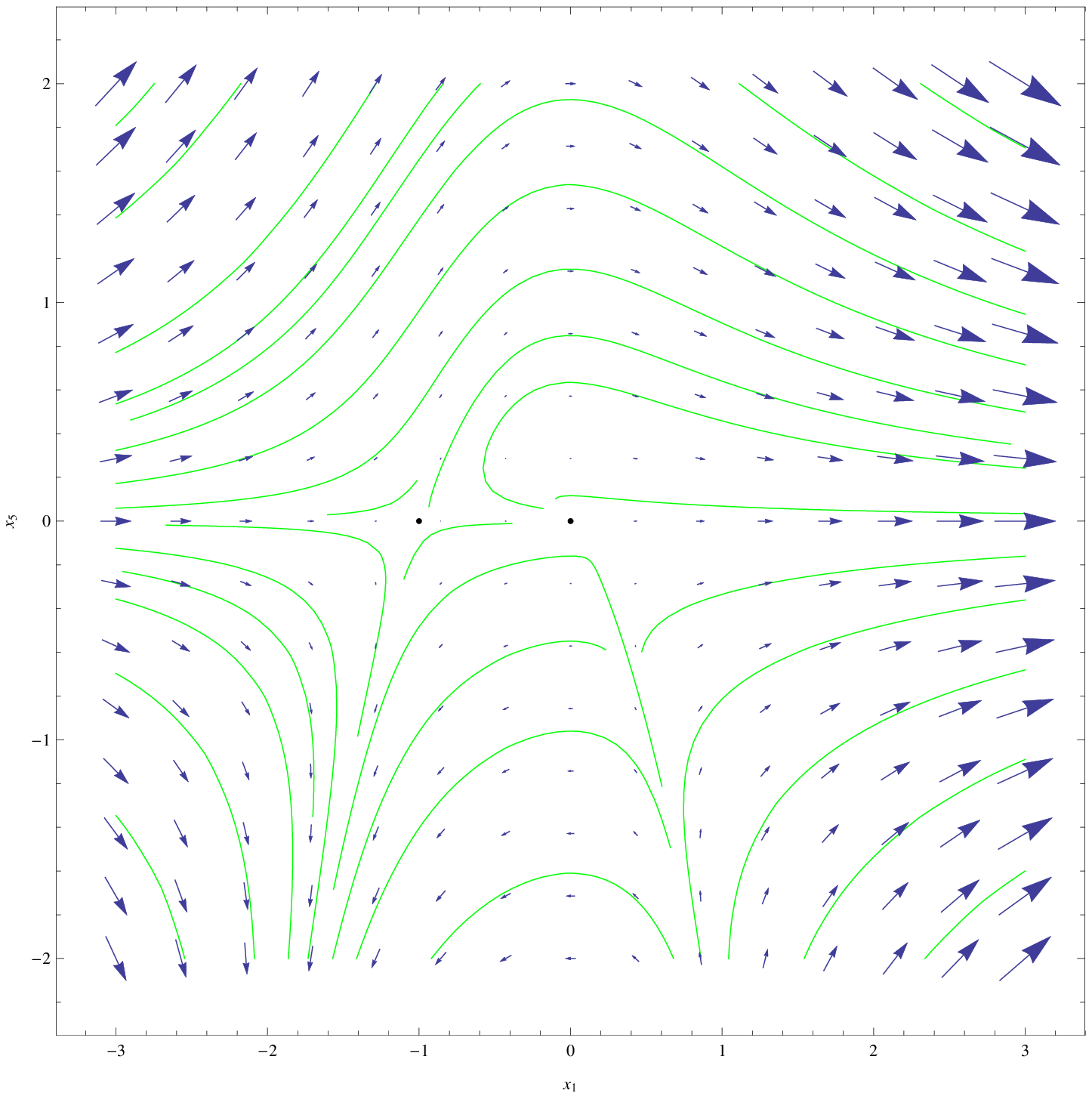}
\caption{Several 2-d intersections of the phase space along the
$x_{1}$ direction, for $c_{1} = -1$, $m=0$, $f_{D} = 1$ and $x_{3}
= 2$. Blue arrows stand for the vector field, green curves for
different solutions and black spots for viable equilibrium
points.} \label{fig:ModelA_m0_PhaseSpace1}
\end{figure}
The values for the parameter $m$ that will concern as were given
earlier, which are $m=0$, $m=-9/2$ and $m=-8$ corresponding to
quasi-de Sitter, matter domination and radiation domination
cosmologies respectively. Finally, concerning the third parameter,
$f_{D}$, no value specification is needed beforehand. It will
prove that all values of $f_{D} > \dfrac{1}{3}$ are capable of
securing at least one stable manifold in the phase space, thus
ensuring some sort of stability. Given that the Friedman
constraint must be fulfilled, a relation between the other two
parameters and $f_{D}$ is given for each of the two distinct
models.

%%%%%%%%%%%%%%%%%%%%%%%%%%%%%%%%%%%%%%%%%%%%%%%%%%%%%%%%%%%%%%%%%%%%%%%%%%%%%%%%%%%%%%%%%%%%%%%%%%%%%%%fridaymorning

\section{The Phase Space of the Model I$\alpha$}

Let us begin by examining the phase space of Model I$\alpha$,
referring to a canonical scalar cosmology. The dynamical system is
composed of Eqs. (\ref{firstODE}, \ref{secondODE}, \ref{thirdODE},
\ref{fourthODEa} and \ref{fifthODE}) and it is subjected to
constraint (\ref{FriedconstrainIa}) and the effective barotropic
index of Eq. (\ref{effeqstate}). The qualitative examination of
the phase space consists mainly of the location and
characterization of the equilibrium points in the phase space.

In the course of this -and the next section- we shall refer to the
vector field defined by differential equations (\ref{firstODE}),
(\ref{secondODE}), (\ref{thirdODE}), (\ref{fourthODEa}) and
(\ref{fifthODE}) as $\vec{V}(x_{1},x_{2},x_{3},x_{4},x_{5})$.

\subsection{Stability and Viability of the Equilibrium Points}

Setting $\vec{V}(x_{1},x_{2},x_{3},x_{4},x_{5}) = 0$, we
analytically derive sixteen critical points, many of them however
come along with complex value in some of their coordinates. This
complexity is mainly attributed to the parameter $m$, out of which
we understand that the system is subject to a number of
bifurcations, mainly due to the parameter $m$ shifting from
positive to negative values.

%\begin{figure}[h!] % "[t!]" placement specifier just for this example
%\begin{floatrow}
%\subfloat[$x_{2}$-$x_{4}$ for $x_{1} = -1$]{
  %  \includegraphics[width=0.27\linewidth]{ModelA_0_x2x4.eps}
  %  \label{fig:2a}}
%\qquad \subfloat[$x_{2}$-$x_{5}$ for $x_{1} = -1$]{
  %  \includegraphics[width=0.27\linewidth]{ModelA_0_x2x5.eps}
   % \label{fig:2b}}
%\qquad \subfloat[$x_{4}$-$x_{5}$ for $x_{1} = -1$ and $x_{2} =
%0$]{
   % \includegraphics[width=0.27\linewidth]{ModelA_0_x4x5.eps}
    %\label{fig:2c}}
%\end{floatrow}

%\caption{2-d intersections of the phase space along the $x_{2}$
%direction, for $c_{1} = -1$, $m=0$, $f_{D} = 1$ and $x_{3} = 2$.
%Blue arrows stand for the vector field, green curves for different
%solutions and black spots for viable equilibrium points.}
%\label{fig:ModelA_m0_PhaseSpace2}
%\end{figure}

If $m > 0$, then all the critical points contain at least one
complex value, so none of them can be an equilibrium point; this
is not really a problem, since $m>0$ does not yield physically
meaningful solutions. On the other hand, if $m = 0$, then none of
the critical points has a complex value and thus they all may be
equilibrium points, however, only six of them exist due to
coincidences. These eight equilibria are characterized by high
degeneracy due to $m = 0$ being the transcritical value in this
bifurcation, and hence the eight equilibrium points exist in a
transitionary state. Finally, if $m < 0$, then six of the
equilibrium points have complex values, hence the remaining ten
are equilibrium points.

The value of $c_{1}$ does not play any role in the number of
equilibrium points existing, neither does the value of $f_{D}$.
However, either of them may alter the stability of the equilibria,
by altering the eigenvalues of the linearized system which is,
\begin{equation} \label{LinIa}
\small
\begin{pmatrix}
\dfrac{d \xi_{1}}{dN} \\ \dfrac{d \xi_{2}}{dN} \\ \dfrac{d
\xi_{3}}{dN} \\ \dfrac{d \xi_{4}}{dN} \\ \dfrac{d \xi_{5}}{dN}
\end{pmatrix}
 =
\begin{pmatrix}
 2 x_{1}^{*}-x_{3}^{*}+3 & 0 & 2-x_{1}^{*} & x_{4}^{*} \left( 2 f_{D} (x_{4}^{*})^2 - c_{1} \right) x_{5}^{*} & \dfrac{1}{2} x_{4}^{*\;2} \left( f_{D} x_{4}^{*\;2} - c_{1} \right) \\
 x_{2}^{*} & x_{1}^{*} -2 x_{3}^{*} + 4 & -2 ( x_{2}^{*} + 2 ) & 0 & 0 \\
 0 & 0 & 8-4 x_{3}^{*} & 0 & 0 \\
 0 & 0 & 0 & -\dfrac{3 \left(-3 c_{1} x_{4}^{*\;2} + 3 f_{D} \left( x_{4}^{*\;2} + 1 \right) x_{4}^{*\;2} + c_{1} \right)}{\left( c_{1}-3 f_{D} x_{4}^{*\;2} \right)^2} & 0 \\
 -x_{5}^{*} & 0 & -2 x_{5}^{*} & 0 & -x_{1}^{*} - 2 x_{3}^{*} + 4
\end{pmatrix}
\begin{pmatrix}
\xi_{1} \\ \xi_{2} \\ \xi_{3} \\ \xi_{4} \\ \xi_{5}
\end{pmatrix} \, ,
\end{equation}
\normalsize where $\{ x_{i}^{*} \}$ indicate the values of the
phase space variables in an equilibrium point and $\{ \xi_{i} \}$
denote small linear perturbations of the phase space variables
around them. It is proved, that in order to ensure structural
stability for almost every equilibrium point, meaning at least one
stable manifold, or at least one eigenvalue with negative real
part in the linearized system, $f_{D}$ must have a lower boundary,
respective to the $c_{1}$ value. Since $c_{1}$ may take only two
possible values, $c_{1} = -1$ and $c_{1} = 1$ respectively,
$f_{D}$ has two lower boundaries arising from the first eigenvalue
of Eqs. (\ref{LinIa}). These are
\begin{itemize}
\item[1] For $c_{1} = -1$, that is in the case of canonical scalar
field, $f_{D}
> \dfrac{1}{3}$. \item[2] For $c_{1} = 1$, that is in the case of
phantom scalar fields, $f_{D} > -\dfrac{1}{3}$.
\end{itemize}

\begin{figure}[h!] % "[t!]" placement specifier just for this example
\includegraphics[width=18pc]{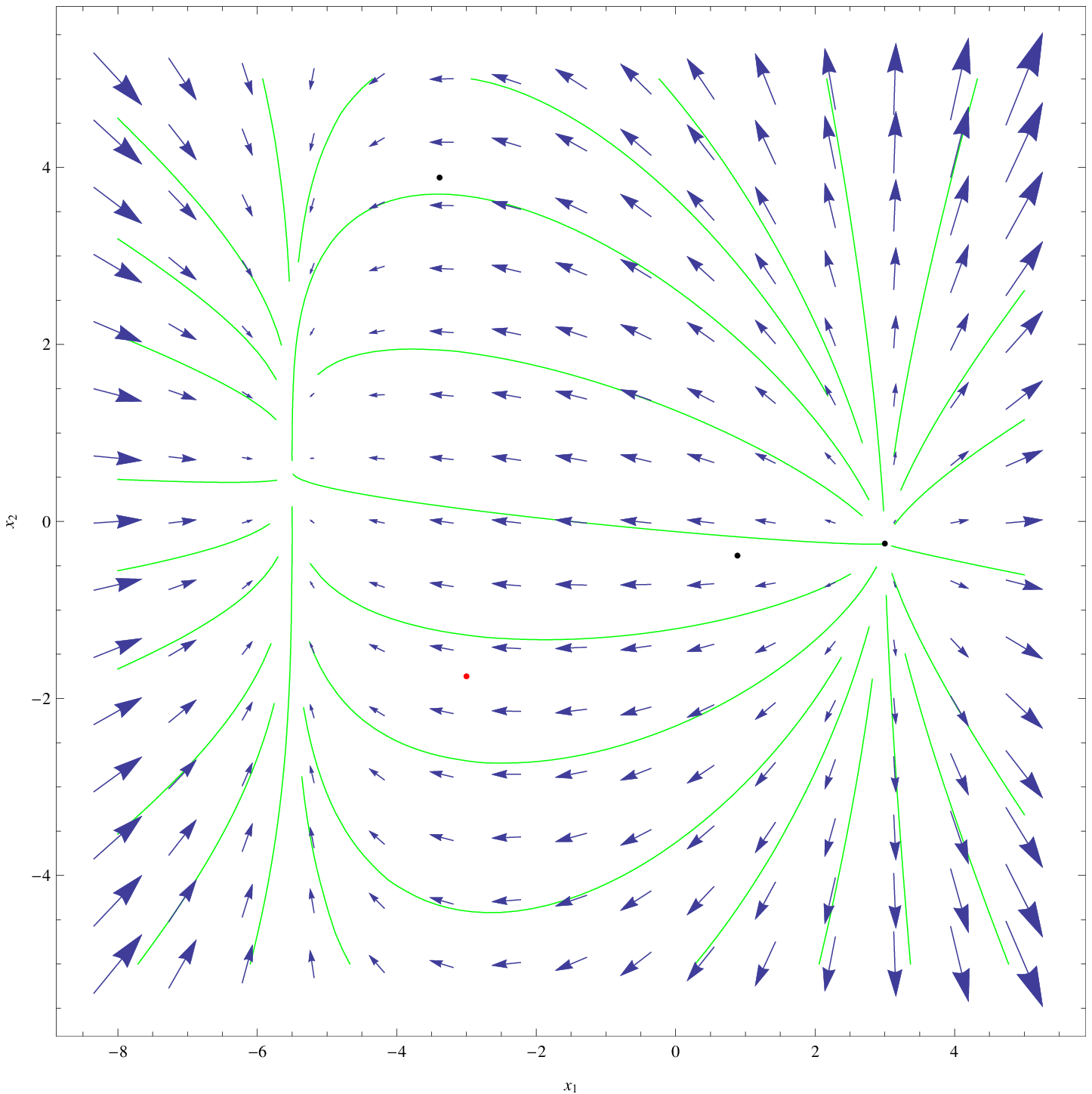}
\includegraphics[width=18pc]{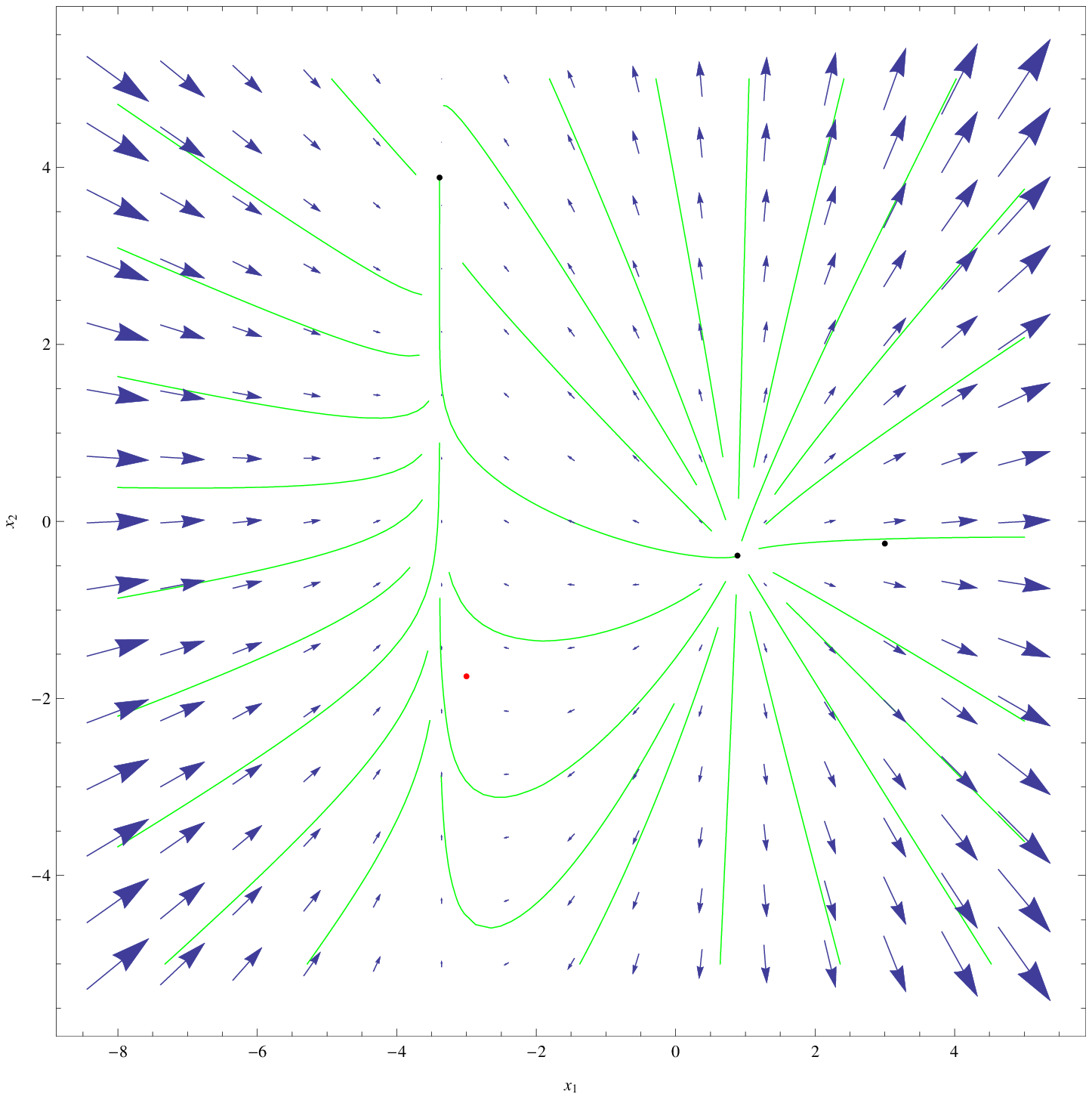}
\includegraphics[width=18pc]{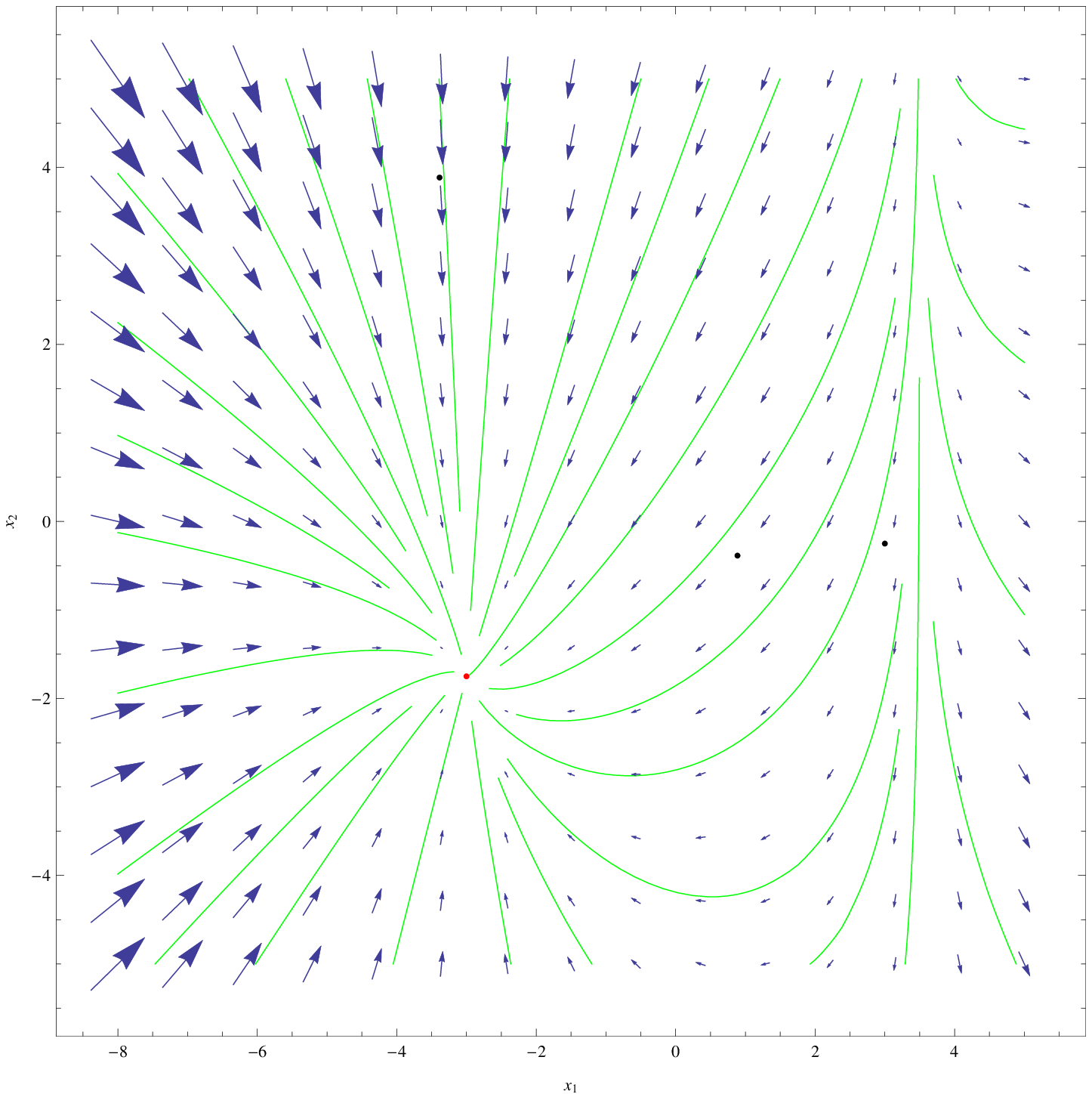}
\caption{Several $x_{1}$-$x_{2}$ intersections of the phase space,
for $c_{1} = -1$, $m = -\dfrac{9}{2}$, $f_{D} = 3$ and $x_{4} =
-1$. The first two plots correspond to $x_{3} = \dfrac{1}{2}$
(viable cosmological solutions), while the third correspond to
$x_{3} = -\dfrac{7}{2}$ (non-viable cosmological solution). Blue
arrows stand for the vector field, green curves for different
solutions, black spots for viable equilibrium points and red spots
for non-viable equilibrium points.}
\label{fig:ModelA_m45_PhaseSpace1}
\end{figure}
Furthermore, the values of $c_{1}$ and $f_{D}$ must also fulfill a
specific relationship, in order to secure the viability of the
majority of the equilibrium points, as the latter is encoded in
the fulfillment of the Friedmann constraint, given in Eq.
(\ref{FriedconstrainIa}), and the effective equation of state, in
Eq. (\ref{effeqstate}). Beginning from the latter, we may easily
rule out any equilibrium point that has a non-matching value for
$x_{3}$. In that case, considering $m \leq 0$, only eight
equilibrium points are deemed viable, regardless of the values of
$c_{1}$ and $f_{D}$.  Moving onto the fulfillment of the former,
we demand that the Friedmann constraint is fulfilled by the eight
equilibrium points at any case and for any possible value of
$c_{1}$, $m$ and $f_{D}$ parameters. Thus we derive the following
equation
\begin{equation} \label{ParameterConstaint}
f_{D} = 1 + 6\dfrac{1 -x_{1}^{*}
-x_{2}^{*}-x_{3}^{*}}{x_{4}^{*\;4} x_{5}^{*}} = \dfrac{\left( 3
c_{1} \sqrt{-2m} + 4 m \right)}{\left( 3 \sqrt{-2m} + 4 m \right)}
\end{equation}

taking into account the ruling out of two equilibria due to the
effective equation of state for $m<0$. From Eq.
(\ref{ParameterConstaint}), we are able to predetermine the value
of the last parameter, $f_{D}$, for the specific values of the
utilized for the other two, $c_{1}$ and $m$. However $c_{1} = 1$
is obviously a pole for $f_{D}$, so in the cases of phantom scalar
fields, $f_{D}$ is considered as a free parameter, chosen as
$f_{D} = \dfrac{1}{2}$ for simplicity and without any lack of
generality.

\subsubsection{Quasi-de Sitter Evolution for $c_{1}=-1$ and $m=0$}

Given $c_{1} = -1$ and $m = 0$, Eq. (\ref{ParameterConstaint}) is
indeterminate, thus $f_{D}$ is indeed a free parameter. We assume
that $f_{D} = 1$ for any necessary calculation, without any lack
of generality. Thus, the six equilibrium points are the following
\begin{equation}
\begin{split}
&P_{1} (0,-1,2,-1,0) \;\;\; \, , \;\;\; P_{2} (0,-1,2,0,0) \;\;\; \, , \;\;\; P_{3} (0,-1,2,1,0) \;\; \, , \\
&P_{4} (-1,0,2,-1,0) \;\;\; \, , \;\;\; P_{5} (-1,0,2,0,0) \;\;\;
\text{and} \;\;\; P_{6}(-1,0,2,1,0) \; \, .
\end{split}
\end{equation}
It is easy to check that all of them fulfill $w_{eff} = -1$ and
the Friedmann constraint, thus all six of them are viable as
cosmological attractor solutions.

Calculating the eigenvalues of the linearized system for these six
equilibria, we come to the conclusion that five out of the six are
structurally stable and more importantly asymptotically unstable,
due to the presence of both positive and negative eigenvalues,
with the sixth being the source of instability.  Furthermore, the
presence of at least one zero eigenvalue in each of them deems
them as irregular and degenerate, a reasonable conclusion due to
the transitional value of $m$ in the bifurcation precess (from
positive $m$'s, where no equilibria exist, to negative $m$'s,
where multiple equilibria arise).

\begin{figure}[h!] % "[t!]" placement specifier just for this example
\includegraphics[width=18pc]{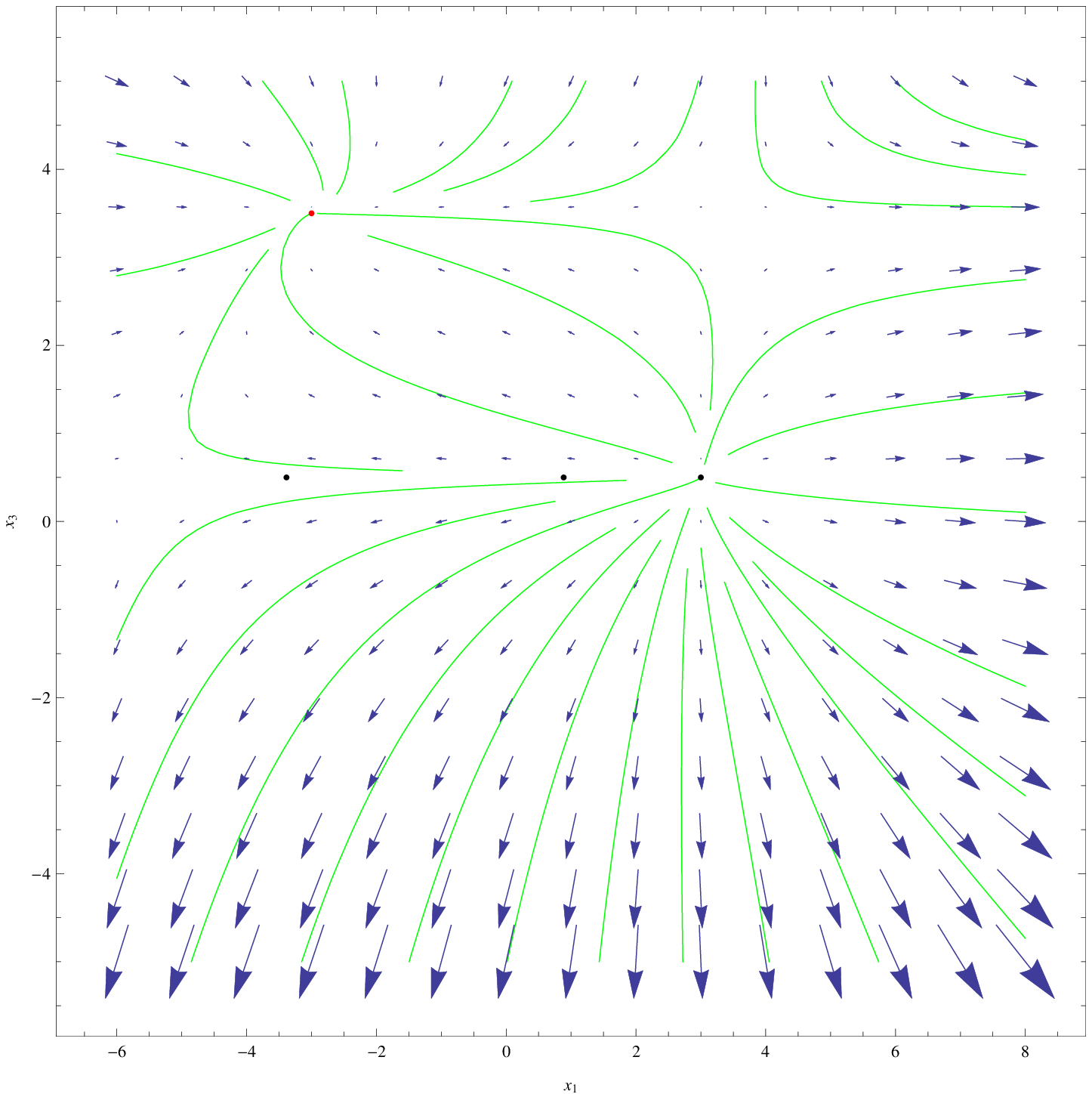}
    \includegraphics[width=18pc]{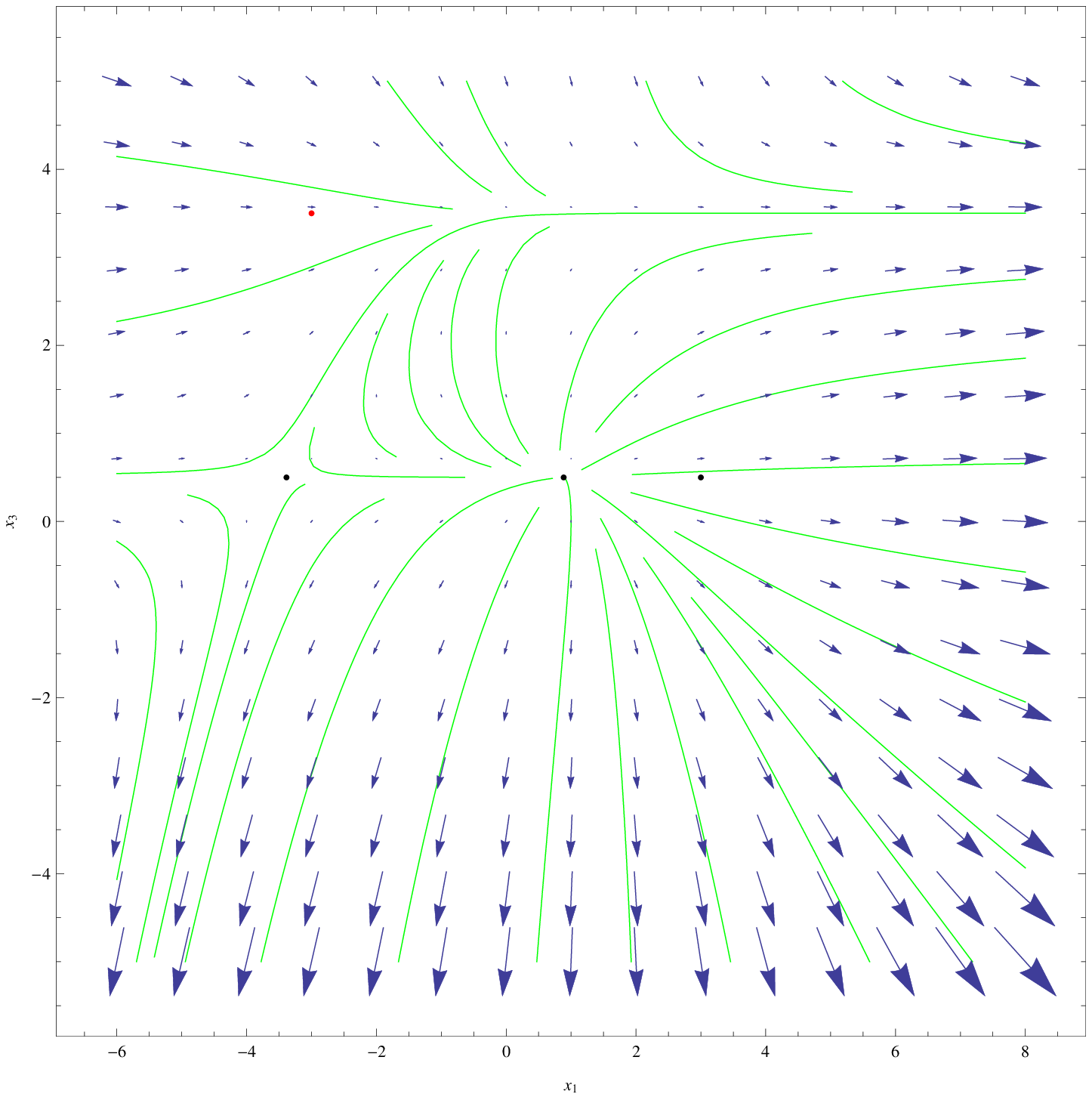}
\caption{2-d intersections of the phase space along the $x_{1}$
direction, for $c_{1} = -1$, $m = -\dfrac{9}{2}$, $f_{D} = 3$ and
$x_{4} = -1$; Both plots correspond to $x_{3} = \dfrac{1}{2}$
(viable cosmological solutions). Blue arrows stand for the vector
field, green curves for different solutions, black spots for
viable equilibrium points and red spots for non-viable equilibrium
points.} \label{fig:ModelA_m45_PhaseSpace2}
\end{figure}
More specifically:
\begin{itemize}

\item[1] The $P_{1}$ and $P_{3}$ equilibrium points have one
stable manifold in the direction of $v_{4} = (0,0,0,1,0)$, and one
unstable in the direction of $v_{1,2} = (-1,1,0,0,0)$; the
remaining three are transitionary. \item[2] The $P_{2}$
equilibrium point has two unstable manifolds in the directions
$v_{1,2} = (-1,1,0,0,0)$ and $v_{4} = (0,0,0,1,0)$; the remaining
three are transitionary. \item[3] The $P_{4}$ and $P_{6}$
equilibrium point have three stable manifolds in the directions
$v_{1} = (1,0,0,0,0)$, $v_{2} = (0,1,0,0,0)$ and $v_{4} =
(0,0,0,1,0)$, and one unstable in the direction of $v_{1,5} =
\Big( \dfrac{1}{2},0,0,0,1 \Big)$. The remaining one is
transitionary. The stability in the directions $v_{1}$ and $v_{2}$
is degenerate, due to equal eigenvalues. \item[4] The $P_{5}$
equilibrium point has two stable manifolds with degenerate
stability, in the directions $v_{1} = (1,0,0,0,0)$ and $v_{2} =
(0,1,0,0,0)$, and two unstable, $v_{4} = (0,0,0,1,0)$ and $v_{5} =
(0,0,0,0,-1)$ and the remaining one is transitionary.
\end{itemize}

In Fig. \ref{fig:ModelA_m0_PhaseSpace1}, we present the behavior
of the phase space variables close to the equilibrium points. It
is easy to see that the attainment of an equilibrium is usually
only along one or two dimensions of the phase space,
%, such as the directions $x_{4} = -1$ and $x_{4} = 1$,
or depends strongly on the initial conditions.
%, such as the directions $x_{3} = 2$ and $x_{5} = 0$.
Generally, the system seems to expand along the directions $x_{1}$
and $x_{5}$, exponentially leading these variables to infinity (or
minus infinity).

The presence of asymptotic instability in the dynamical system,
after some quasi-de Sitter attractors are reached, is particularly
physically appealing. This is due to the fact that inflationary
attractors are reached, and then the phase space structure of the
$k$-Essence $f(R)$ gravity reaches some unstable manifolds
(certain directions in the phase space), leading to the conclusion
that the inflationary attractors become destabilized. This can be
viewed as an inherent mechanism for graceful exit from inflation
in the $k$-Essence $f(R)$ gravity theory. Thus combining the
present results with those of Ref. \cite{Nojiri:2019dqc} which
indicated compatibility of $k$-Essence $f(R)$ gravity theory with
the latest Planck data, this makes the theory particularly useful
for describing inflationary dynamics.

\subsubsection{Matter-dominated Era: The Case $c_{1}=-1$ and $m=-\dfrac{9}{2}$}

Let us focus in the case of having canonical scalar fields and
matter domination cosmology, which is achieved by choosing $c_{1}
= -1$ and $m = -\dfrac{9}{2}$. In effect, Eq.
(\ref{ParameterConstaint}) yields $f_{D} = 3$ so that the
Friedmann constraint will be satisfied at least for those
equilibria yielding $w_{eff} = 0$. The ten equilibrium points in
this case read,
\begin{equation}
\begin{split}
&P_{1}\Big( 3,-\dfrac{1}{4},\dfrac{1}{2},-1,-\dfrac{27}{4} \Big) \;\; \, , \;\; P_{2} \Big( -3,-\dfrac{7}{4},\dfrac{7}{2},-1,-\dfrac{27}{4} \Big) \;\;\; \, , \;\;\; P_{3} \Big( 3,-\dfrac{1}{4},\dfrac{1}{2},1,-\dfrac{27}{4} \Big) \;\; \, , \;\; P_{4} \Big( -3,-\dfrac{7}{4},\dfrac{7}{2},1,-\dfrac{27}{4} \Big) \;\; \, , \\
&P_{5} \Big( -\dfrac{5+\sqrt{73}}{4},\dfrac{7+\sqrt{73}}{4},\dfrac{1}{2},-1,0 \Big) \;\; \, , \;\; P_{6} \Big( -\dfrac{5+\sqrt{73}}{4},\dfrac{7+\sqrt{73}}{4},\dfrac{1}{2},0,0 \Big) \;\; \, , \;\; P_{7} \Big( -\dfrac{5+\sqrt{73}}{4},\dfrac{7+\sqrt{73}}{4},\dfrac{1}{2},1,0 \Big) \;\; \, , \\
&P_{8} \Big(
-\dfrac{5-\sqrt{73}}{4},\dfrac{7-\sqrt{73}}{4},\dfrac{1}{2},-1,0
\Big) \;\; \, , \;\; P_{9} \Big(
-\dfrac{5-\sqrt{73}}{4},\dfrac{7-\sqrt{73}}{4},\dfrac{1}{2},0,0
\Big) \;\; \text{and} \;\; P_{10} \Big(
-\dfrac{5-\sqrt{73}}{4},\dfrac{7-\sqrt{73}}{4},\dfrac{1}{2},1,0
\Big) \;\; \, .
\end{split}
\end{equation}
We can easily check that points $P_{2}$ and $P_{4}$ yield $w_{eff}
= -2$ and do not satisfy the Friedmann constraint of Eq.
(\ref{FriedconstrainIa}). As a result, they correspond to
non-viable cosmologies, however all other equilibrium points yield
$w_{eff} = 0$ and satisfy the Friedmann constraint.

In order to account for the stability of the equilibrium points,
we calculate the eigenvalues of the linearized system (Eq.
(\ref{LinIa})). Again, all points but one turn to be structurally
stable and asymptotically unstable, with at least one unstable
manifold. The tenth point is proved unstable. More analytically,
\begin{itemize}
\item[1]Points $P_{1}$ and $P_{3}$ have two stable and three
unstable manifolds; two of the three unstable manifolds are
degenerate due to the equality of the corresponding eigenvalues.
\item[2] Non-viable points $P_{2}$ and $P_{4}$ have four stable
manifolds and one unstable; two of the four stable manifolds are
degenerate as the equality of the corresponding eigenvalues
suggest. \item[3] Points $P_{5}$ and $P_{7}$ have three stable
manifolds, in the directions of $v_{1,2} = (-1,1,0,0,0)$, $v_{2} =
(0,1,0,0,0)$ and $v_{4} = (0,0,0,1,0)$, and two unstable. \item[4]
Point $P_{6}$ has two stable manifolds, in the directions of
$v_{1,2} = (-1,1,0,0,0)$ and $v_{2} = (0,1,0,0,0)$, and three
unstable manifolds. \item[5] Points $P_{8}$ and $P_{10}$ have one
stable manifold, in the direction of $v_{4} = (0,0,0,1,0)$, and
four unstable manifolds. \item[6] Point $P_{9}$ has five unstable
manifolds, being an unstable node.
\end{itemize}
This behavior can partly be seen in Fig.
\ref{fig:ModelA_m45_PhaseSpace2} where we present the phase space
structure in terms of some  of the phase space variables.

\subsubsection{Radiation Dominated Era: The Case $c_{1}=-1$ and $m=-8$}

Now let us consider the radiation domination cosmologies, in which
case $m=-8$ and let us also investigate the case $c_{1} = -1$
which corresponds to canonical scalar fields. For  $c_{1}=-1$ and
$m=-8$, Eq. (\ref{ParameterConstaint}) yields $f_{D} =
\dfrac{11}{5}$ so that the Friedmann constraint will be satisfied
at least for those equilibria yielding $w_{eff} = \dfrac{1}{3}$.
The ten equilibrium points in this case read,
\begin{equation}
\begin{split}
&P_{1}( 4,0,0,-1,-15 ) \;\;\; \, , \;\;\; P_{2} ( -4,-2,4,-1,-15 ) \;\;\; \, , \;\;\; P_{3} ( 4,0,0,1,-15 ) \;\;\; \, , \;\;\; P_{4} ( -4,-2,4,1,-15 ) \;\; \, , \\
&P_{5} ( -4,5,0,-1,0 ) \;\;\; \, , \;\;\; P_{6} ( -4,5,0,0,0 ) \;\;\; \, , \;\;\; P_{7} ( -4,5,0,1,0 ) \;\; \, , \\
&P_{8} ( 1,0,0,0,-1,0 ) \;\;\; \, , \;\;\; P_{9} ( 1,0,0,0,0,0 )
\;\;\; \text{and} \;\;\; P_{10} ( 1,0,0,0,1,0 ) \, .
\end{split}
\end{equation}
We can easily check that points $P_{2}$ and $P_{4}$ yield $w_{eff}
= -\dfrac{7}{3}$ and do not satisfy the Friedmann constraint of
Eq. (\ref{FriedconstrainIa}). As a result, they correspond to
non-viable cosmologies. All other equilibrium points yield
$w_{eff} = \dfrac{1}{3}$ and satisfy the Friedmann constraint.

As for the stability of the equilibrium points, the eigenvalues of
the linearized system (Eq. (\ref{LinIa})) for each of them reveal
a complex and unstable nature, similar to the previous case. All
points are accompanied by at least one unstable manifold and those
providing the greater stability (four stable and one unstable
manifolds) are the non-viable two, $P_{2}$ and $P_{4}$. More
specifically,
\begin{itemize}
\item[1]Points $P_{1}$ and $P_{3}$ have two stable and three
unstable manifolds; two of the three unstable manifolds are
degenerate due to the equality of the corresponding eigenvalues.
\item[2] Points $P_{2}$ and $P_{4}$ have four stable and one
unstable manifolds; two of the four stable manifolds are
degenerate as the equality of the corresponding eigenvalues
suggest. \item[3] Points $P_{5}$ and $P_{7}$ have two stable
manifolds, in the directions $v_{1,2} = (-1,1,0,0,0)$ and $v_{4} =
(0,0,0,1,0)$, and two degenerate unstable manifolds; they also
have a transitionary one. \item[4] Point $P_{6}$ has one stable
manifold, in the direction $v_{1,2} = (-1,1,0,0,0)$ and three
unstable, two of which are degenerate since their corresponding
eigenvalues are equal; it also has a transitionary manifold
corresponding to a zero eigenvalue. \item[5] Points $P_{8}$ and
$P_{10}$ have one stable manifold, in the direction $v_{4} =
(0,0,0,1,0)$, and four unstable; two of the four unstable are
degenerate, due to the equality of the corresponding eigenvalues.
\item[6] Point $P_{9}$ has five unstable manifolds, four of whom
are pairwise degenerate, being a degenerate unstable node.
\end{itemize}
\begin{figure}[h!] % "[t!]" placement specifier just for this example
 \includegraphics[width=18pc]{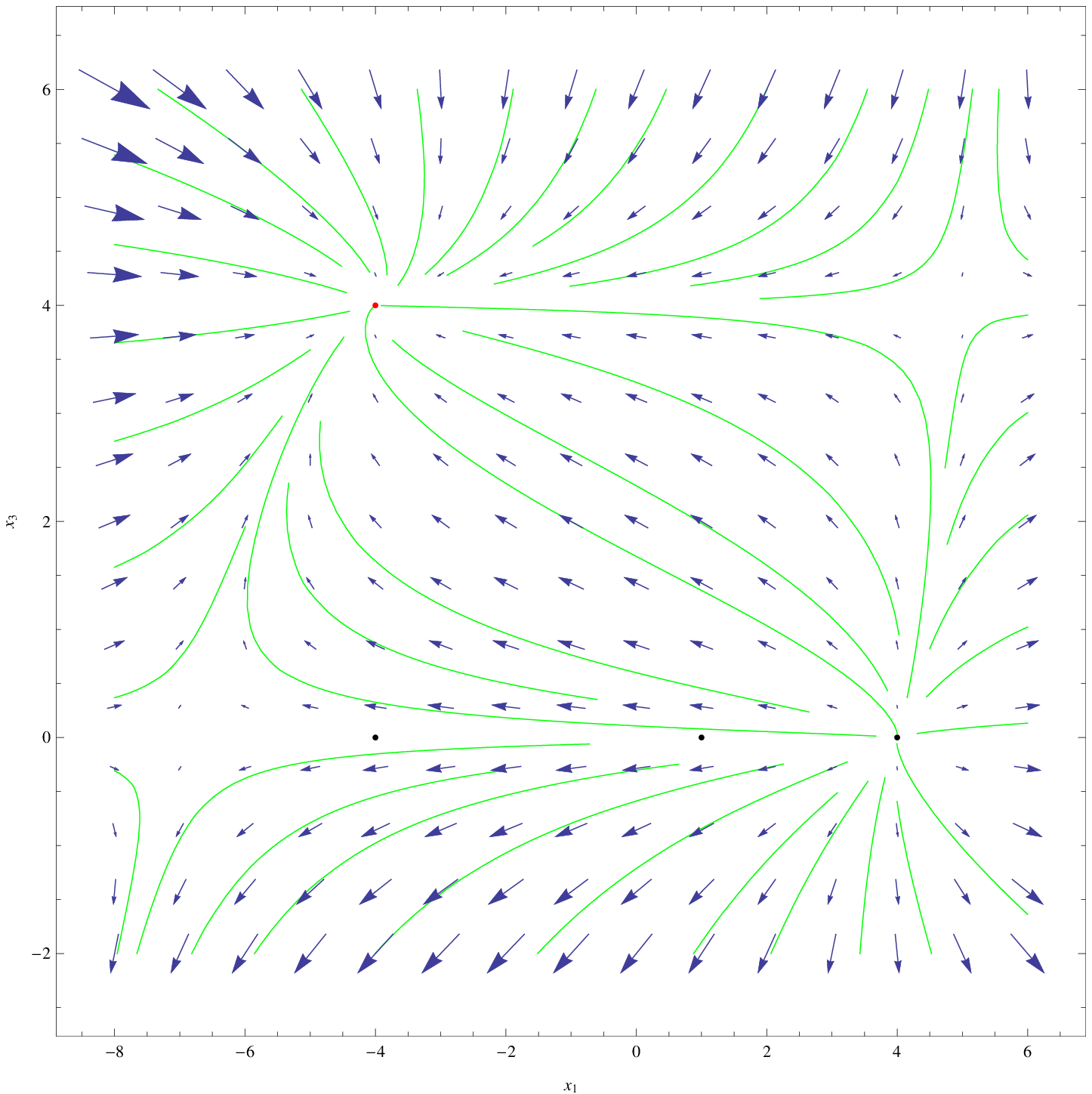}
    \includegraphics[width=18pc]{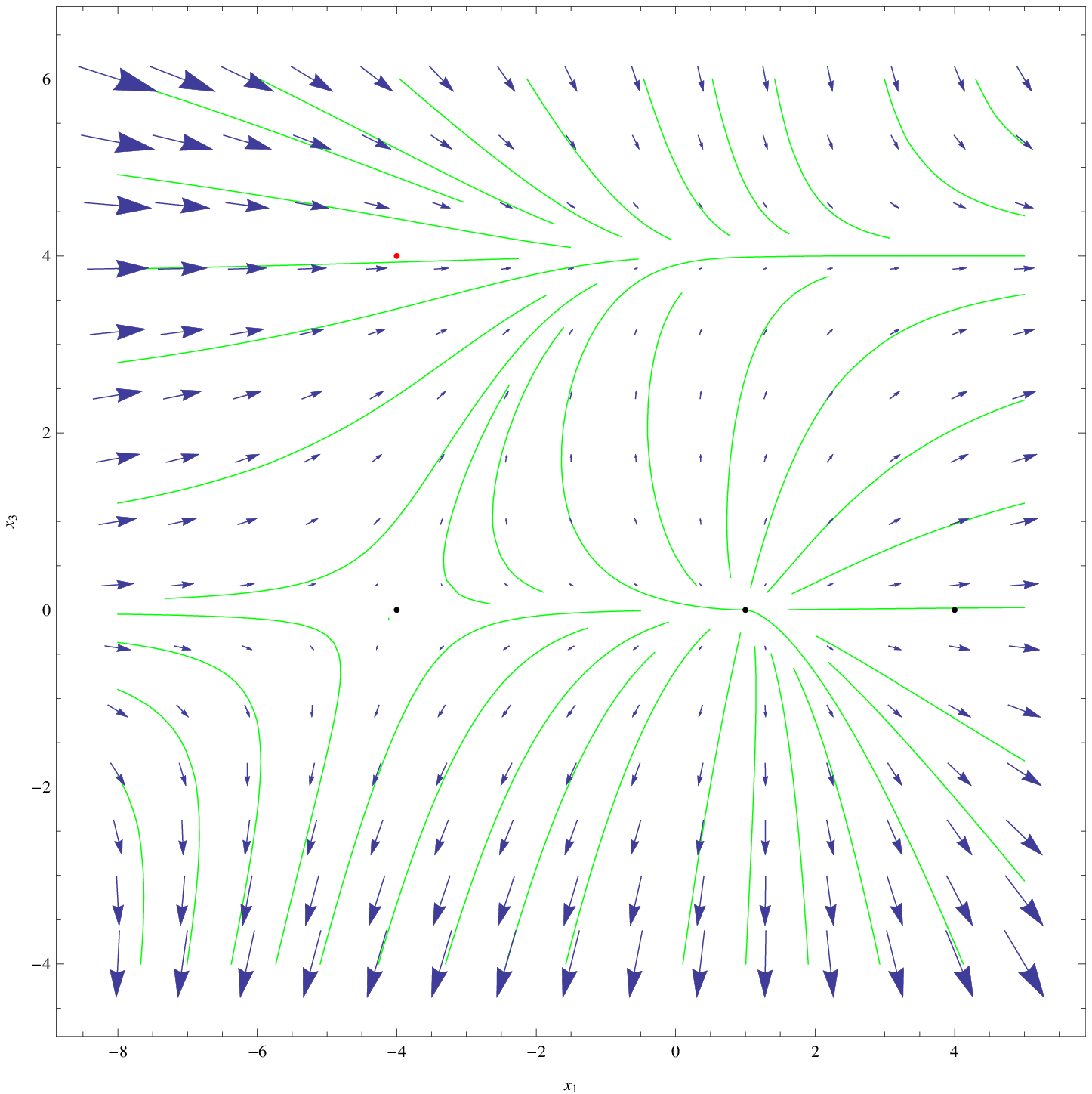}
    \includegraphics[width=18pc]{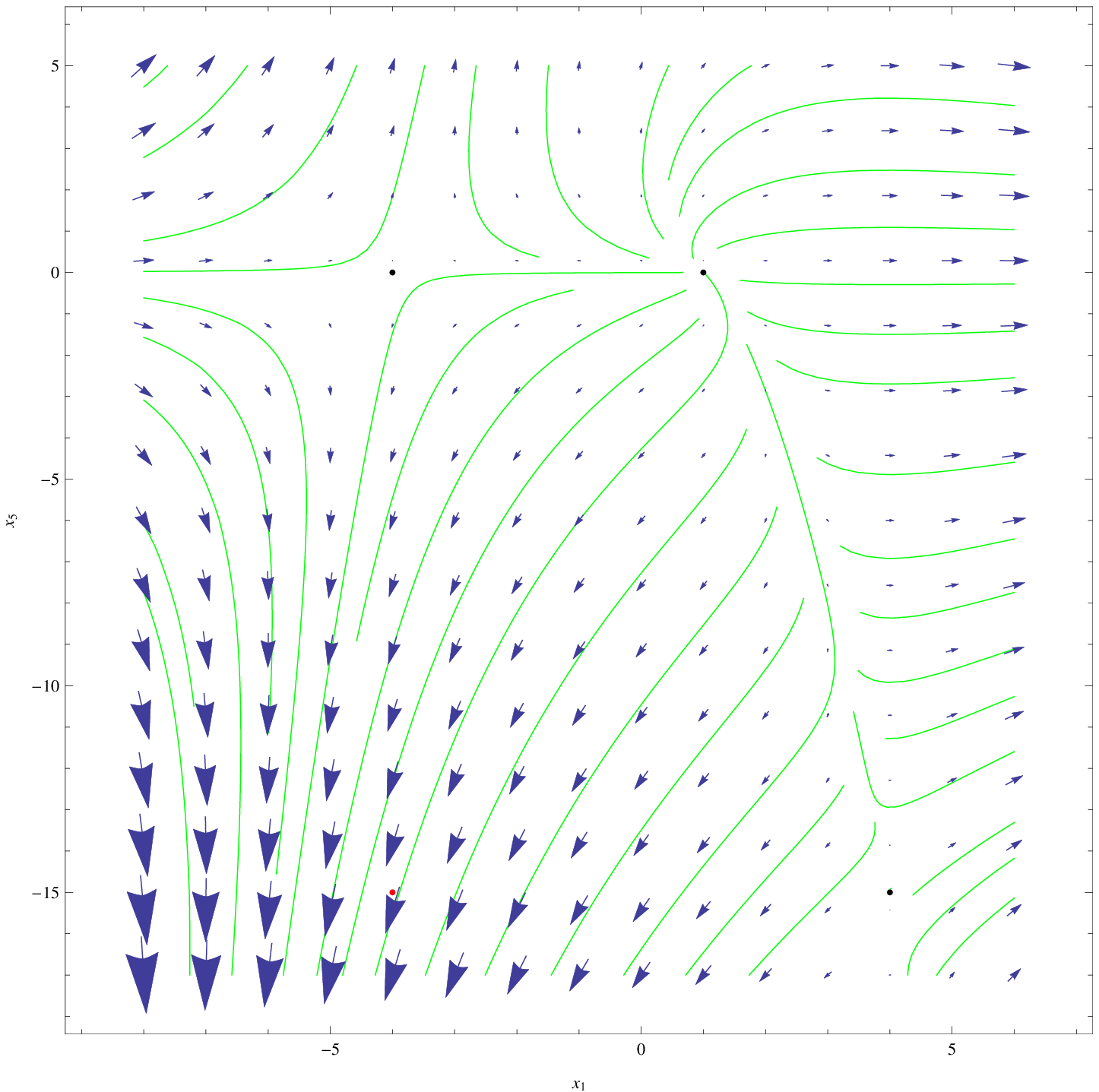}
\caption{2-d intersections of the phase space along the $x_{1}$
direction, for $c_{1} = -1$, $m = -8$, $f_{D} = 3$. Blue arrows
stand for the vector field, green curves for different solutions,
black spots for viable equilibrium points and red spots for
non-viable equilibrium points.} \label{fig:ModelA_m8_PhaseSpace2}
\end{figure}
In Fig. \ref{fig:ModelA_m8_PhaseSpace2} we present the behavior of
some phase space variables, for $c_{1} = -1$, $m = -8$, $f_{D} =
3$. It can be seen that the phase space has attractors, which
eventually become destabilized.

\subsubsection{Quasi-de Sitter Evolution with Phantom Scalar Fields: The Case $c_{1}=1$ and $m=0$}

Let us now consider quasi-de Sitter cosmologies, accompanied by
phantom scalar fields. In this case $c_{1} = 1$, the parameter
$f_{D}$ is not determined by Eq. (\ref{ParameterConstaint}) and
chosen as $f_{D} = \dfrac{1}{2}$; this value is sustained for all
phantom scalar field cases. The six equilibrium points of the
system when $c_{1} = 1$ and $m = 0$ are,
\begin{equation}
\begin{split}
&P_{1} (0,-1,2,-1,0) \;\;\; \, , \;\;\; P_{2} (0,-1,2,0,0) \;\;\; \, , \;\;\; P_{3} (0,-1,2,1,0) \;\; \, , \\
&P_{4} (-1,0,2,-1,0) \;\;\; \, , \;\;\; P_{5} (-1,0,2,0,0) \;\;\;
\text{and} \;\;\; P_{6} (-1,0,2,1,0) \, .
\end{split}
\end{equation}
The Friedmann constraint of Eq. (\ref{FriedconstrainIa}) is
generally satisfied and so is the effective equation of state,
yielding $w_{eff} = -1$, in contrast to the phantom case, where
some equilibria where unphysical.

\begin{figure}[h!] % "[t!]" placement specifier just for this example
\centering
    \includegraphics[width=18pc]{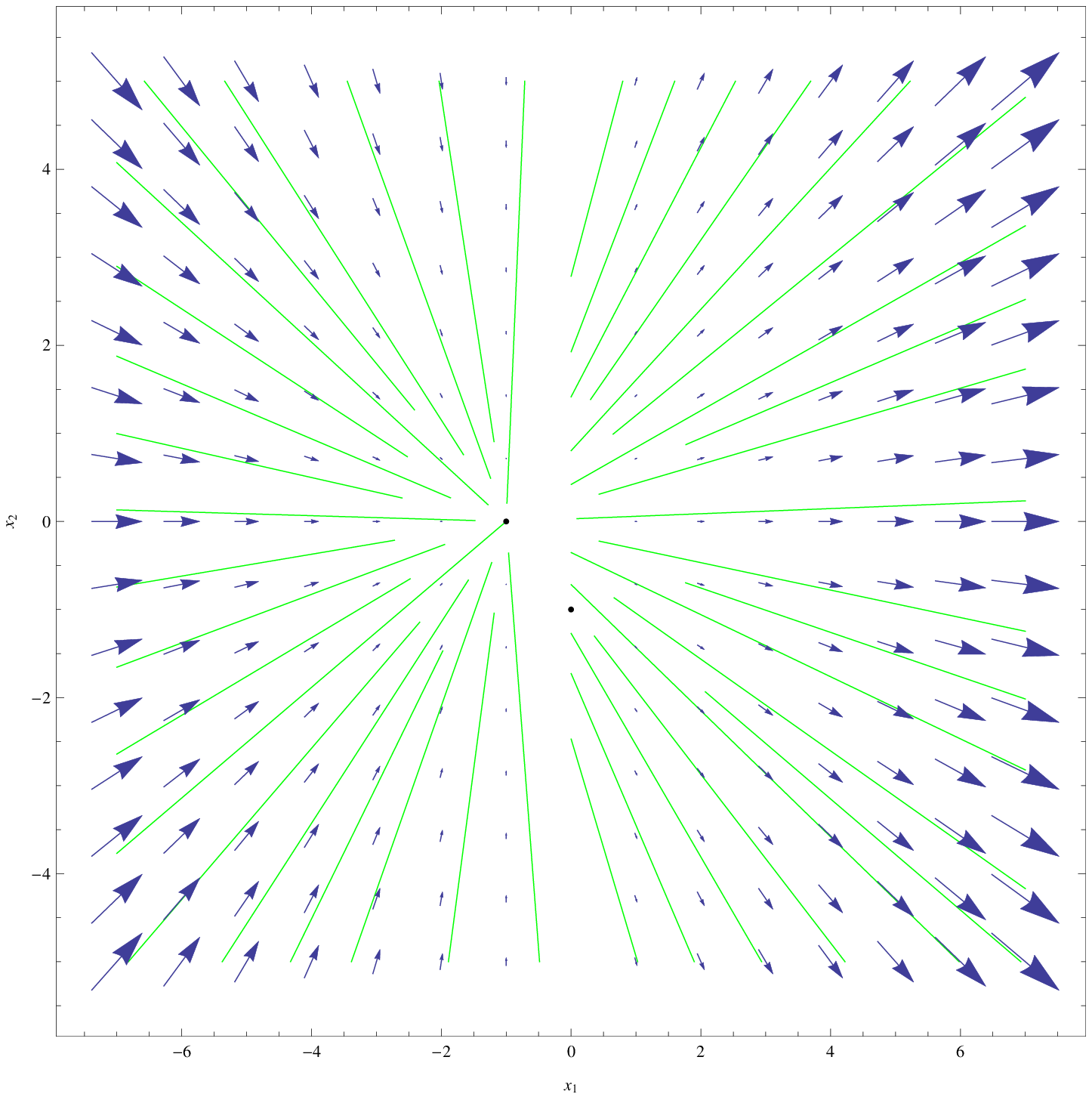}
    \includegraphics[width=18pc]{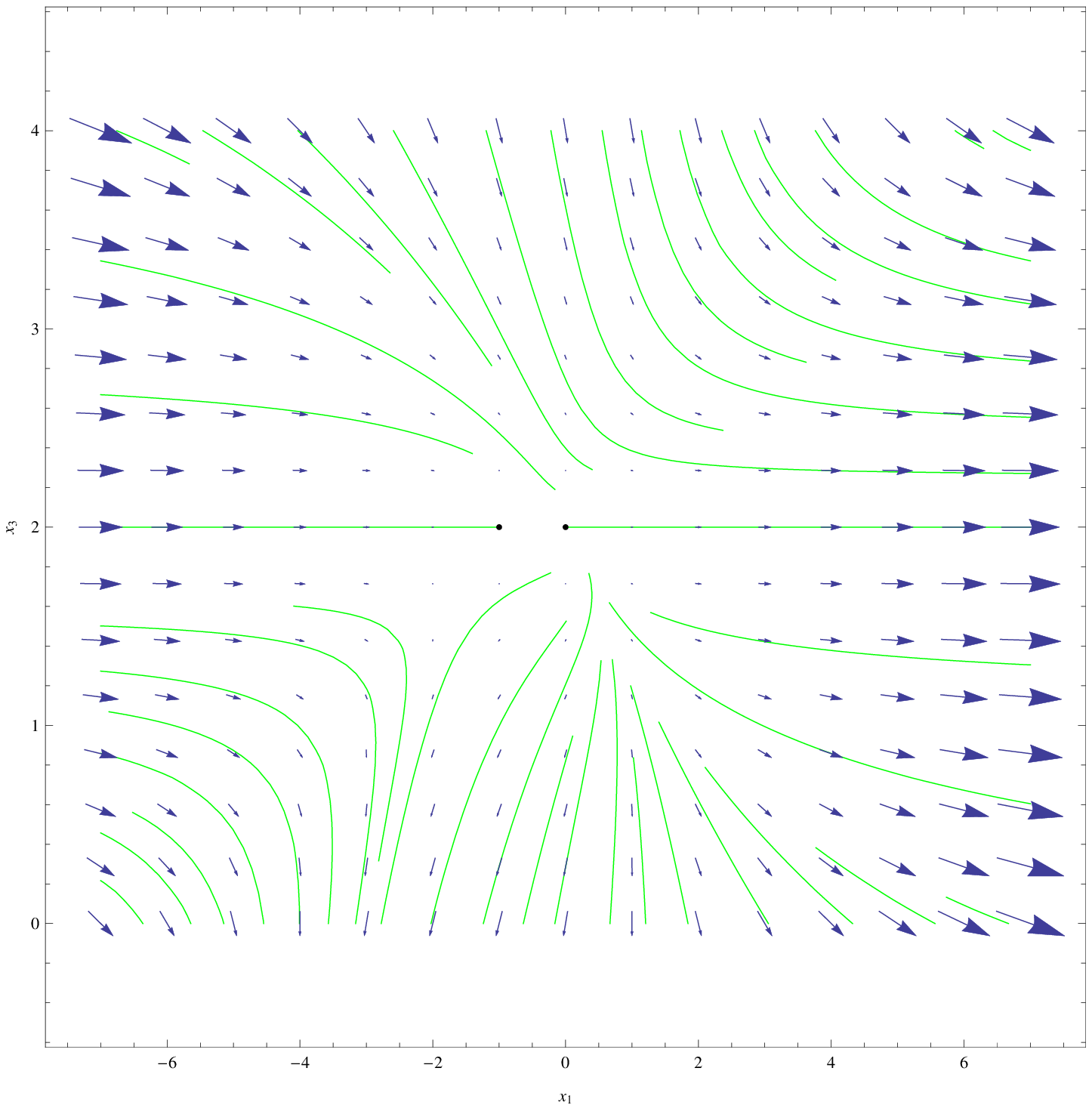}
    \includegraphics[width=18pc]{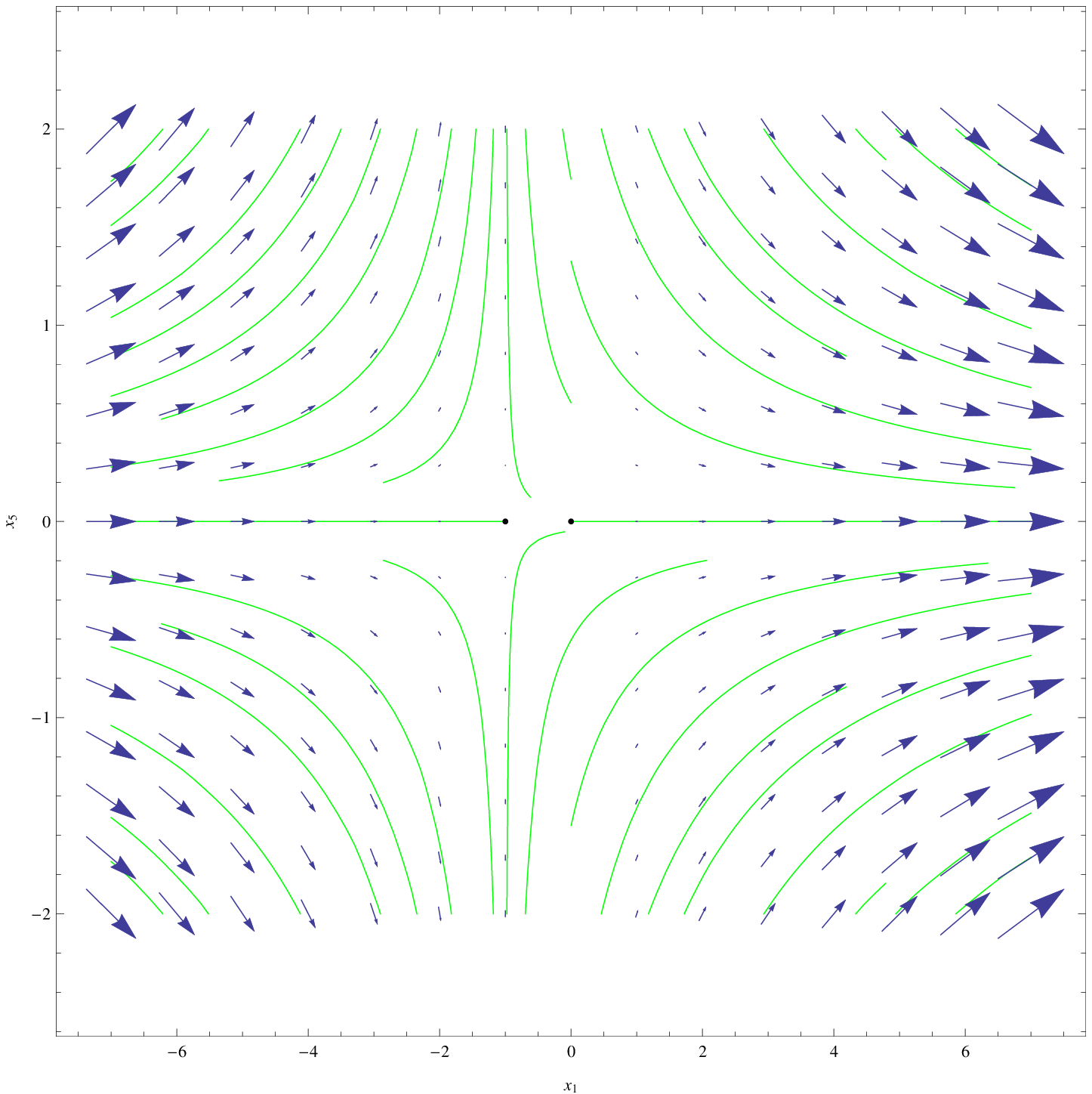}
\caption{ 2-d intersections of the phase space along the $x_{1}$
direction, for $c_{1} = 1$, $m=0$, $f_{D} = \dfrac{1}{2}$ and
$x_{3} = 2$ (left and right plot), $x_{4} = -1$. Blue arrows stand
for the vector field, green curves for different solutions and
black spots for viable equilibrium points.}
\label{fig:ModelB_m0_PhaseSpace1}
\end{figure}

Much similarly to the respective phantom scalar case studied
earlier, the majority of the manifolds around the equilibrium
points are transitionary, since $m = 0$ is the indicative value
for the bifurcation and the turning point for the sign of the
eigenvalues. Furthermore, no clear stability is established for
any point, with at least one unstable manifold being present in
every one. Specifically,
\begin{itemize}
\item [1] Points $P_{1}$, $P_{2}$ and $P_{3}$ have one stable
manifold, in the direction of $v_{4} = (0,0,0,1,0)$, and one
unstable, in the direction of $v_{1,2} = (-1,1,0,0,0)$; the
remaining three are transitionary due to the zero corresponding
eigenvalues. \item[2] Points $P_{4}$, $P_{5}$ and $P_{6}$ have
three stable manifolds, in the directions $v_{1} = (1,0,0,0,0)$ ,
$v_{2} = (0,1,0,0,0)$ and $v_{4} = (0,0,0,1,0)$, and one unstable
manifold; the remaining one is accompanied by a zero eigenvalue,
thus being transitionary.
\end{itemize}
The result is intriguing, since it seems that the canonical
$k$-Essence $f(R)$ gravity theory is more physically appealing in
comparison to the phantom scalar $k$-Essence $f(R)$ gravity. This
result is of particular interest since it is aligned with the
results of Ref. \cite{Nojiri:2019dqc} indicating the same result,
that the canonical scalar $k$-Essence $f(R)$ gravity is compatible
with the Planck data, without extreme fine-tuning. In Fig.
\ref{fig:ModelB_m0_PhaseSpace1} we plot the behavior of several
phase space variables for $c_{1} = 1$, $m=0$, $f_{D} =
\dfrac{1}{2}$. The instability we mentioned above is apparent in
all plots.

\subsubsection{Matter-dominated era with Phantom fields: $c_{1}=1$ and $m=-\dfrac{9}{2}$}

Let us now consider matter dominated cosmologies with phantom
scalar fields, so in this case $c_{1} = 1$, $m = -\dfrac{9}{2}$
and $f_{D} = \dfrac{1}{2}$. The corresponding equilibrium points
become ten and are the following,
\begin{equation}
\begin{split}
&P_{1}\Big( 3,-\dfrac{1}{4},\dfrac{1}{2},-1,18 \Big) \;\; \, , \;\; P_{2} \Big( -3,-\dfrac{7}{4},\dfrac{7}{2},-1,18 \Big) \;\;\; \, , \;\;\; P_{3} \Big( 3,-\dfrac{1}{4},\dfrac{1}{2},1,18 \Big) \;\; \, , \;\; P_{4} \Big( -3,-\dfrac{7}{4},\dfrac{7}{2},1,18 \Big) \;\; \, , \\
&P_{5} \Big( -\dfrac{5+\sqrt{73}}{4},\dfrac{7+\sqrt{73}}{4},\dfrac{1}{2},-1,0 \Big) \;\; \, , \;\; P_{6} \Big( -\dfrac{5+\sqrt{73}}{4},\dfrac{7+\sqrt{73}}{4},\dfrac{1}{2},0,0 \Big) \;\; \, , \;\; P_{7} \Big( -\dfrac{5+\sqrt{73}}{4},\dfrac{7+\sqrt{73}}{4},\dfrac{1}{2},1,0 \Big) \;\; \, , \\
&P_{8} \Big(
-\dfrac{5-\sqrt{73}}{4},\dfrac{7-\sqrt{73}}{4},\dfrac{1}{2},-1,0
\Big) \;\; \, , \;\; P_{9} \Big(
-\dfrac{5-\sqrt{73}}{4},\dfrac{7-\sqrt{73}}{4},\dfrac{1}{2},0,0
\Big) \;\; \text{and} \;\; P_{10} \Big(
-\dfrac{5-\sqrt{73}}{4},\dfrac{7-\sqrt{73}}{4},\dfrac{1}{2},1,0
\Big) \;\; \, .
\end{split}
\end{equation}
Of all of them, only five are viable, satisfying both the
Friedmann constraint, Eq. (\ref{FriedconstrainIa}), and the
effective equation of state, Eq. (\ref{effeqstate}). Points
$P_{1}$ and $P_{3}$ satisfy the constraint, but yield $w_{eff} =
-\dfrac{7}{3}$, while point $P_{7}$ satisfies the equation of
state, but not the Friedman constraint. Finally points $P_{2}$ and
$P_{4}$ satisfy neither constraint.

\begin{figure}[h!] % "[t!]" placement specifier just for this example
\centering
\includegraphics[width=18pc]{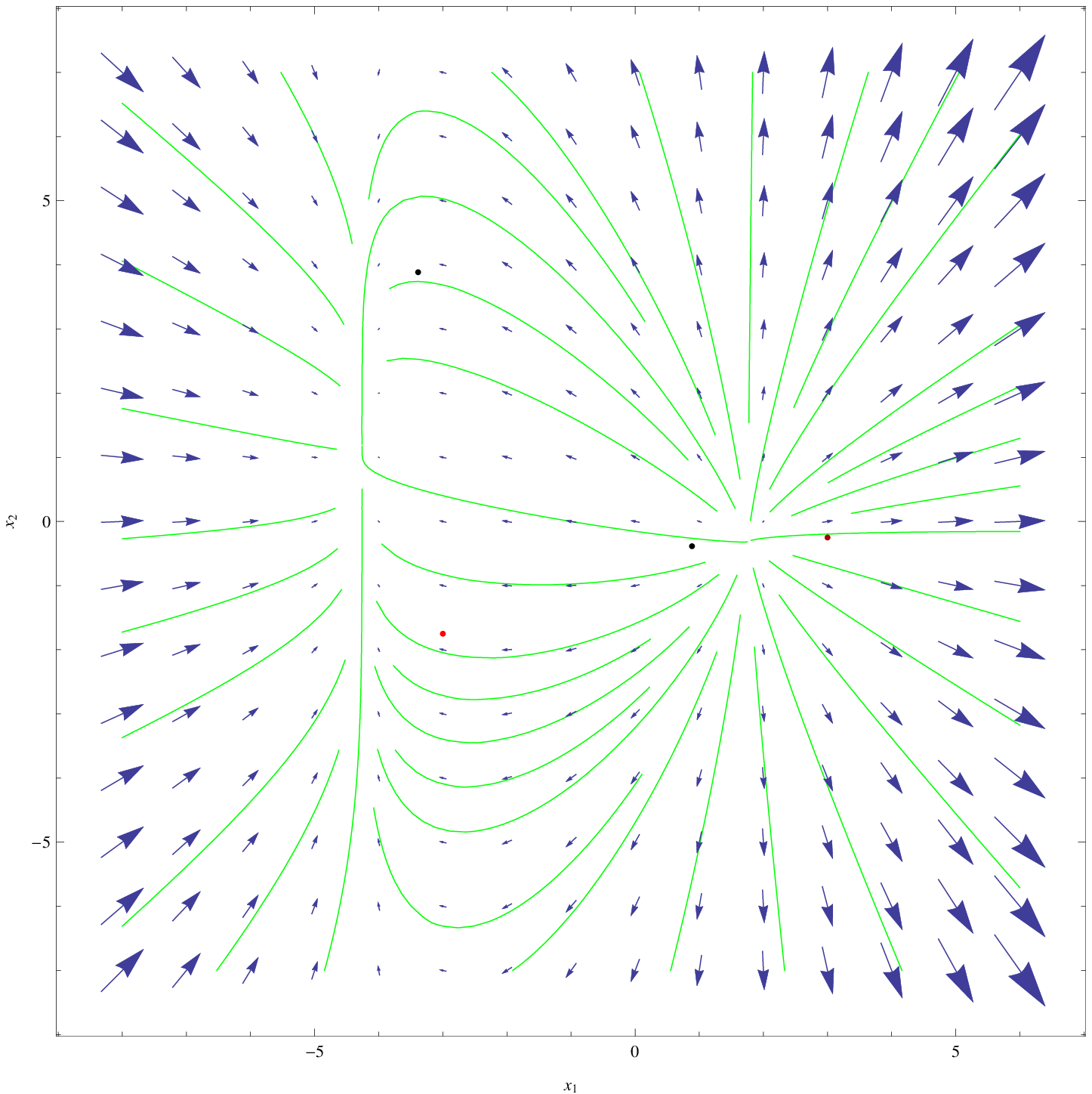}
\includegraphics[width=18pc]{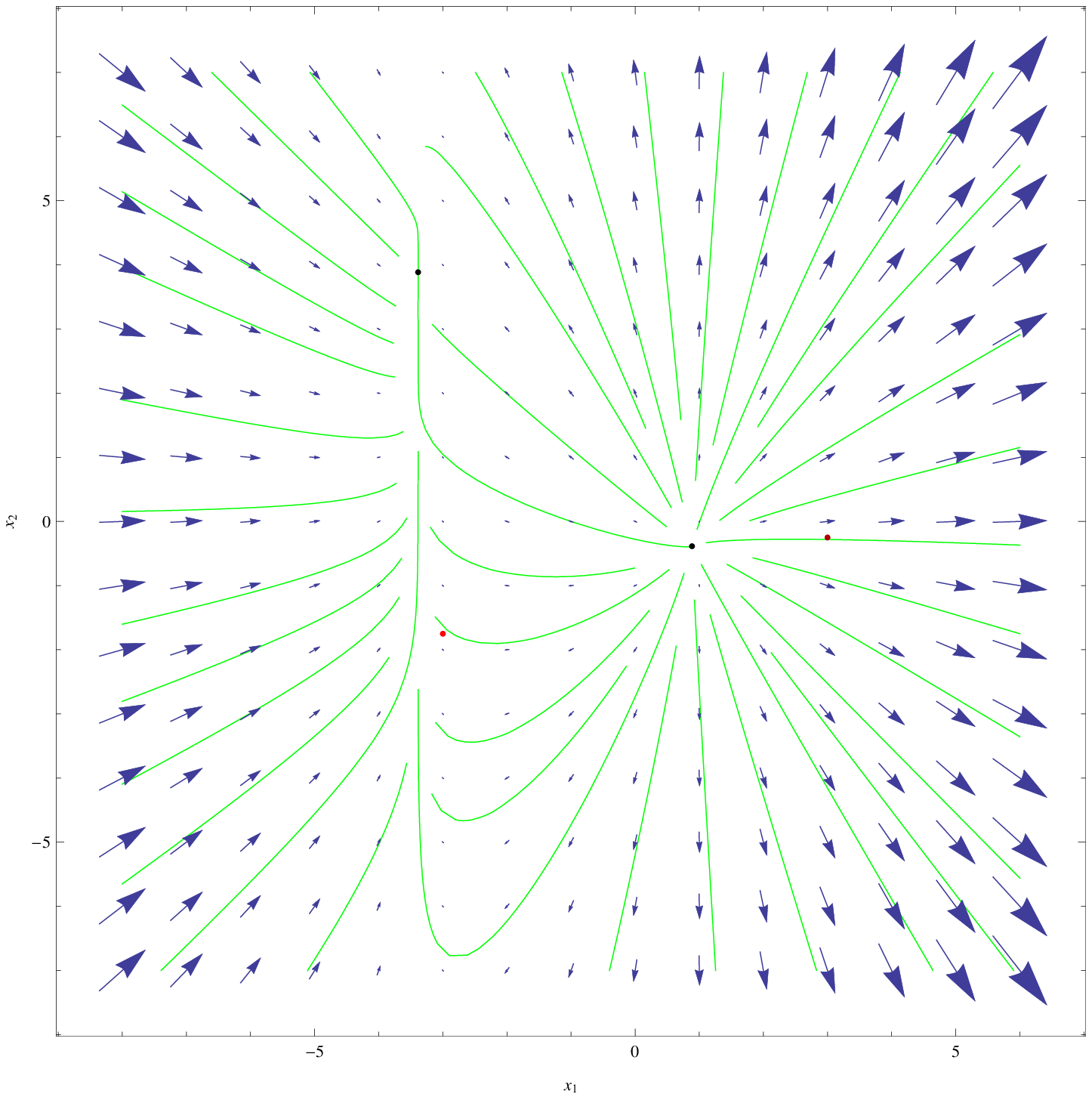}
\caption{ 2-d intersections of the phase space along the $x_{1}$
direction, for $c_{1} = 1$, $m = -\dfrac{9}{2}$, $f_{D} =
\dfrac{1}{2}$ and $x_{3} = \dfrac{1}{2}$ (corresponding to
physically viable cosmological solutions). Blue arrows stand for
the vector field, green curves for different solutions, black
spots for viable equilibrium points and crimson and red spots for
non-viable equilibrium points.} \label{fig:ModelB_m45_PhaseSpace1}
\end{figure}

Concerning their stability, we again derive the eigenvalues of the
linearized system of Eq. (\ref{LinIa}), and our analysis indicates
that,
\begin{itemize}
\item[1] Points $P_{1}$ and $P_{3}$ have two stable and three
unstable manifolds; two of the latter are degenerate due to the
equality of the corresponding eigenvalues. \item[2] Points $P_{2}$
and $P_{4}$ have four stable and one unstable manifolds; two of
the former are degenerate as the equality of the corresponding
eigenvalues suggest. \item[3] Points $P_{5}$ and $P_{7}$ have
three stable manifolds, in the directions $v_{1,2} =
(-1,1,0,0,0)$, $v_{2} = (0,1,0,0,0)$ and $v_{4} = (0,0,0,1,0)$,
and two unstable. \item[4] Point $P_{6}$ has four stable
manifolds, in the directions $v_{1,2} = (-1,1,0,0,0)$, $v_{2} =
(0,1,0,0,0)$, $v_{4} = (0,0,0,1,0)$ and $v_{5} = (0,0,0,0,1)$, and
one unstable. \item[5] Points $P_{8}$, $P_{9}$ and $P_{10}$ have
one stable manifold, in the direction $v_{4} = (0,0,0,1,0)$, and
four unstable manifolds.
\end{itemize}
The structure of the space is depicted in Fig.
\ref{fig:ModelB_m45_PhaseSpace1} for some of the phase space
variables, for $c_{1} = 1$, $m = -\dfrac{9}{2}$, $f_{D} =
\dfrac{1}{2}$. In this case too, structural instabilities occur in
the phase space, and this is a generic feature of the phantom
scalar field $k$-Essence gravity.

\subsubsection{Radiation Dominated Era with Phantom Scalar Fields: The Case $c_{1}=1$ and $m=-8$}

Let us finally consider the case where phantom scalar fields are
considered for radiation dominated cosmologies, in which case
$c_{1} = 1$, $m = -8$ and $f_{D} = \dfrac{1}{2}$. Our analysis
indicates that the following ten equilibrium points exist,
\begin{equation}
\begin{split}
&P_{1}( 4,0,0,-1,32 ) \;\;\; \, , \;\;\; P_{2} ( -4,-2,4,-1,32 ) \;\;\; \, , \;\;\; P_{3} ( 4,0,0,1,32 ) \;\;\; \, , \;\;\; P_{4} ( -4,-2,4,1,32 ) \;\; \, , \\
&P_{5} ( -4,5,0,-1,0 ) \;\;\; \, , \;\;\; P_{6} ( -4,5,0,0,0 ) \;\;\; \, , \;\;\; P_{7} ( -4,5,0,1,0 ) \;\; \, , \\
&P_{8} ( 1,0,0,0,-1,0 ) \;\;\; \, , \;\;\; P_{9} ( 1,0,0,0,0,0 )
\;\;\; \text{and} \;\;\; P_{10} ( 1,0,0,0,1,0 ) \, .
\end{split}
\end{equation}
Of these ten, the six satisfy both the Friedmann constraint of Eq.
(\ref{FriedconstrainIa}) and the effective equation of state (Eq.
(\ref{effeqstate})), yielding $w_{eff} = 0$, meaning that the
phantom fields do not affect relativistic matter either. However,
though points $P_{1}$ and $P_{3}$ yield $w_{eff} = -\dfrac{1}{3}$,
they do not satisfy the constraint. Finally points $P_{2}$ and
$P_{4}$ do satisfy neither of the constraints.

\begin{figure}[h!] % "[t!]" placement specifier just for this example
\centering
    \includegraphics[width=18pc]{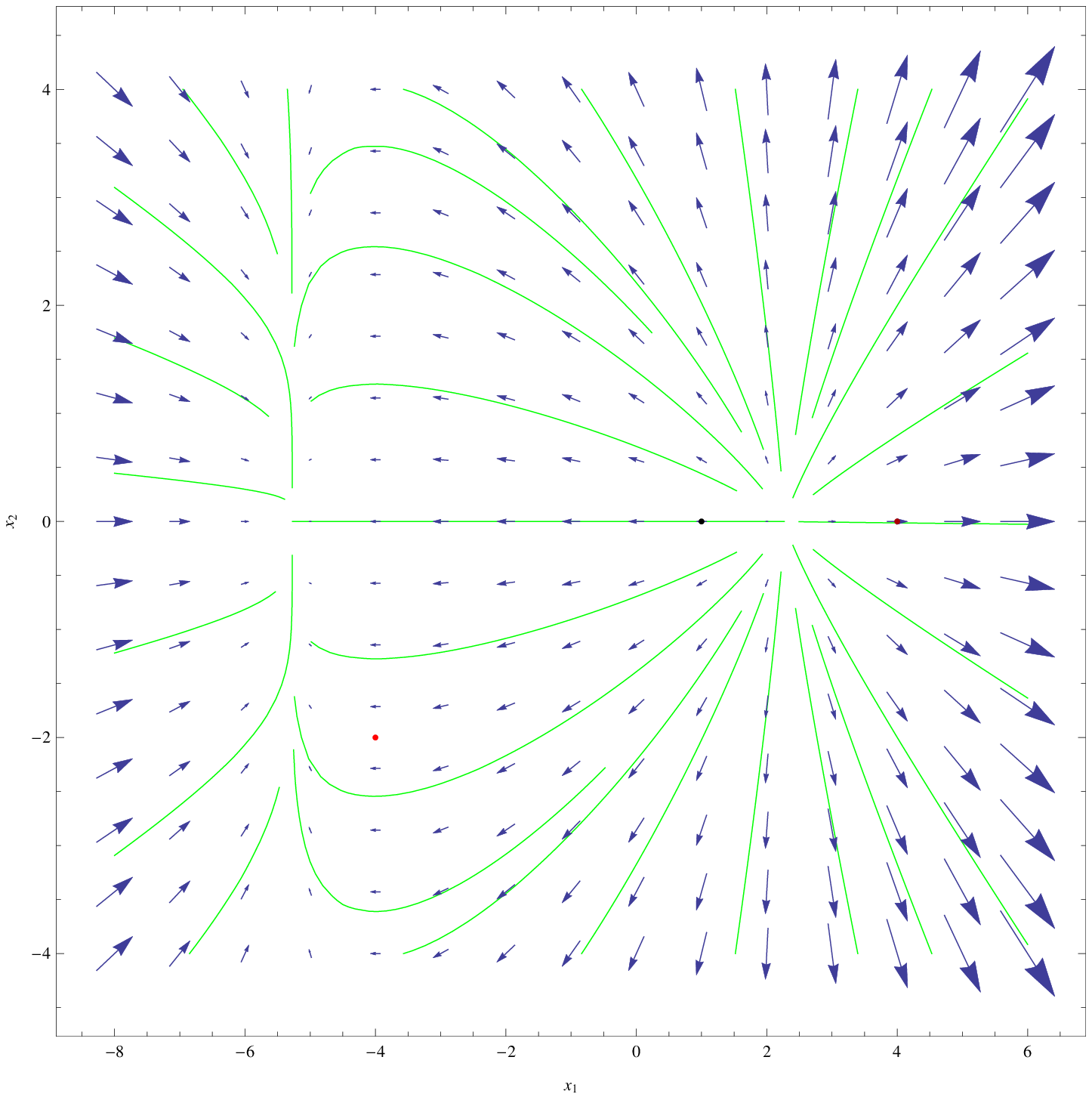}
    \includegraphics[width=18pc]{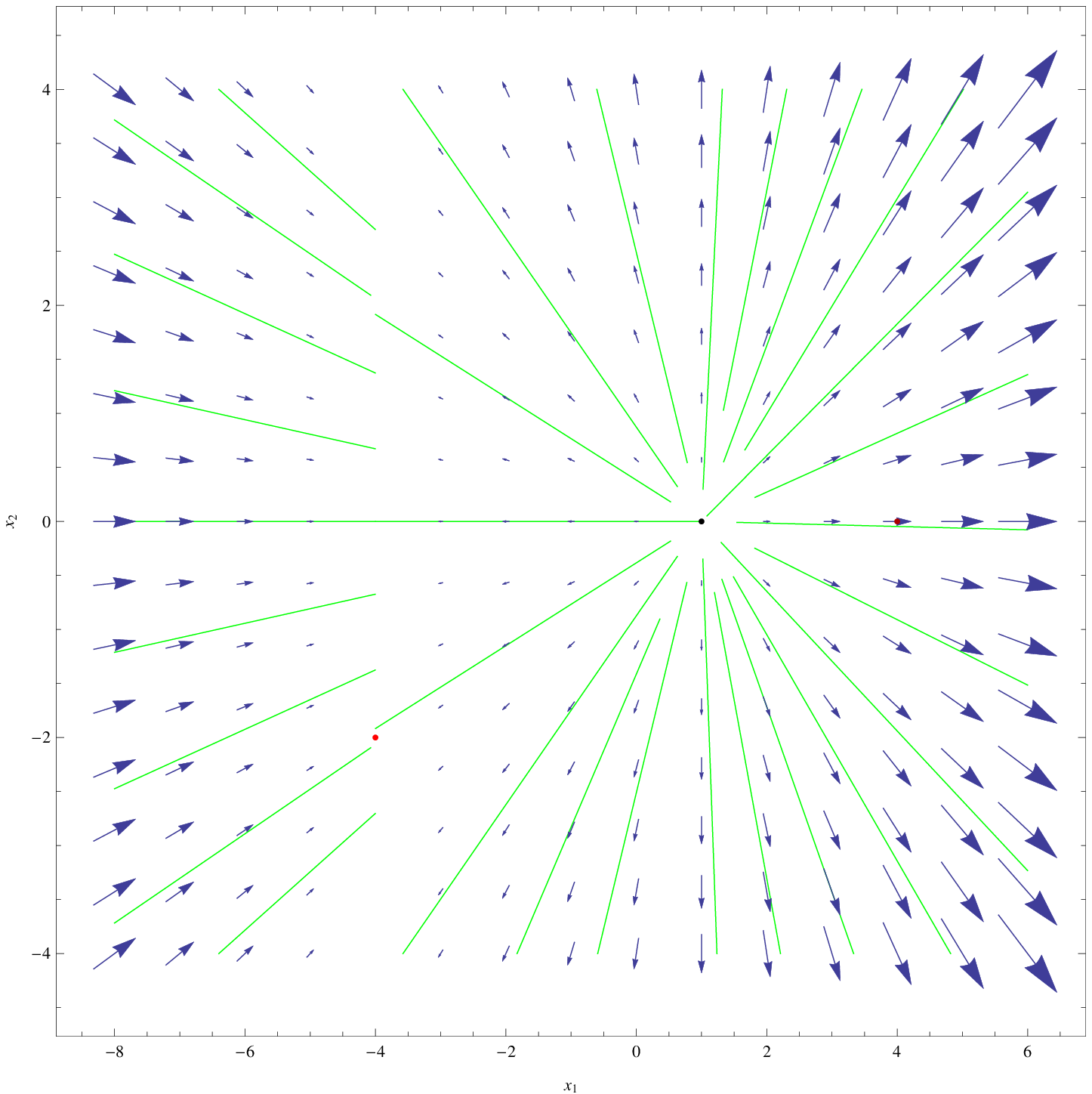}
    \includegraphics[width=18pc]{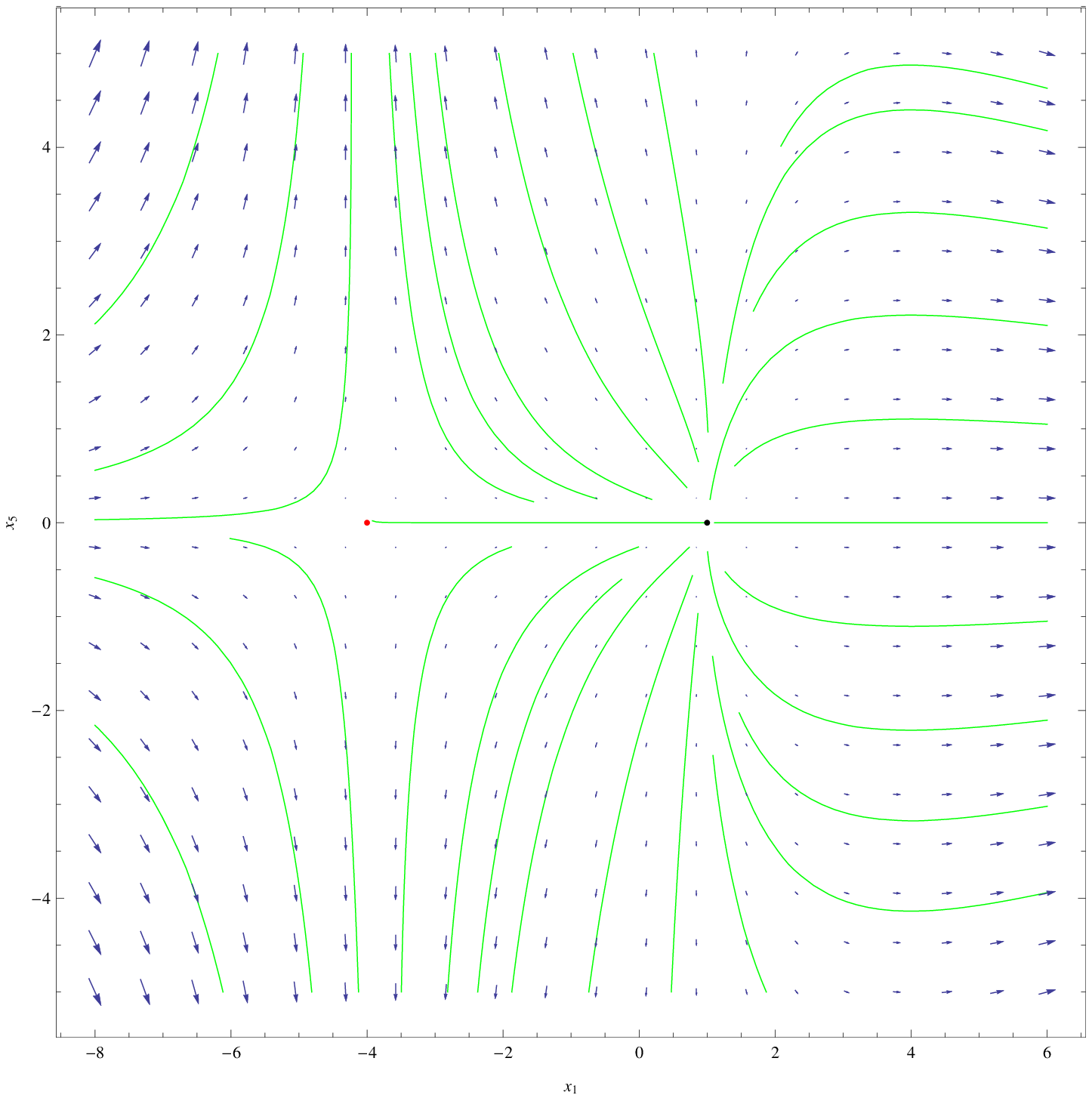}
\caption{ Several $x_{1}$-$x_{2}$ intersections of the phase
space, for $c_{1} = 1$, $m = -8$ and $f_{D} = \dfrac{1}{2}$. Blue
arrows stand for the vector field, green curves for different
solutions, black spots for viable equilibrium points and crimson
and red spots for non-viable equilibrium points.}
\label{fig:ModelB_m8_PhaseSpace1}
\end{figure}

As for their stability, we may again check the eigenvalues of the
linearized system of Eq. (\ref{LinIa}), and obtain the following
results:
\begin{itemize}
\item[1] Points $P_{1}$ and $P_{3}$ have two stable and three
unstable manifolds; two of the latter are degenerate since the
corresponding eigenvalues are equal. \item[2] Points $P_{2}$ and
$P_{4}$ have four stable and one unstable manifolds; of the
former, two are degenerate as their equal eigenvalues suggest.
\item[3] Points $P_{5}$, $P_{6}$ and $P_{7}$ have two stable
manifolds, in the directions $v_{1,2} = (-1,1,0,0,0)$ and $v_{4} =
(0,0,0,1,0)$, and two degenerate unstable manifolds, due to the
equality of their eigenvalues; the remaining one is transitionary,
as the zero eigenvalue suggest. \item[4] Points $P_{8}$, $P_{9}$
and $P_{10}$ have one stable manifold, in the direction $v_{4} =
(0,0,0,1,0)$ and four unstable; two of the latter are proved
degenerate due to the equality of the corresponding eigenvalues.
\end{itemize}
The above results can be clearly seen in Fig.
\ref{fig:ModelB_m8_PhaseSpace1}, for $c_{1} = 1$, $m = -8$, $f_{D}
= \dfrac{1}{2}$.

%%%%%%%%%%%%%%%%%%%%%%%%%%%%%%%%%%%%%%%%%%%%%%%%%%%%%%%%%%%%%%%%%%%%%%%%%%%%%%%%%%%%%%%fridaynight

\subsection{A Possible $2-d$ Attractor}

Very important information about a dynamical system arise from the
divergence of its vector field,
$\vec{V}(x_{1},x_{2},x_{3},x_{4},x_{5})$, which for the model at
hand is equal to,
\begin{equation} \label{flowIa}
\vec{\nabla} \vec{V} = 13 + 2 x_{1} - 9 x_{3} + \dfrac{1}{f_{D}}
\Big( -1 - \dfrac{c_{1} - 3 f_{D}}{c_{1} - 3 f_{D} x_{4}^2} +
\dfrac{2 c_{1} (c_{1} - 3 f_{D} )}{( c_{1} - 3 f_{D} x_{4}^2
)^{2}} \Big)\, .
\end{equation}
Generally, a dynamical system is explosive if $\vec{\nabla}
\vec{V} > 0$, conservative if $\vec{\nabla} \vec{V} = 0$, or
dissipative if $\vec{\nabla} \vec{V} < 0$, meaning that
supervolumes of initial values are increasing, non-changing, or
decreasing over time, respectively. In our case, the sign of the
divergence of the flow changes, which means that the system is
neither explosive, neither conservative, nor dissipative, but
rather a mixture of all these depending on the phase space where
the flow operates on the initial values.

According to the Poincar\'{e}-Bendixon theorem, the change of sign
of the flow of a dynamical system indicates the existence of an
attractor or a repeller in the phase, such as a stable or unstable
limit cycle. The case of our dynamical system is similar, since
the flow, $\vec{\nabla} \vec{V}$ turns zero along a specific
three-dimensional curve, which is defined as follows,
\begin{equation} \label{limcycle}
x_{4}^{2} = \dfrac{ 3 f_{D} + c_{1} \big( 2 f_{D} (13 + 2 x_{1} -
9 x_{3}) - 3 \big) - \sqrt{ (3 f_{D} - c_{1})\big[ 3 f_{D} + c_{1}
\big( 3 f_{D} - 9 + c_{1} (13 + 2 x_{1} - 9 x_{3}) \big) \big] }
}{6 f_{D} \big( f_{D}(13 + 2 x_{1} - 9 x_{3}) - 1 \big) } \, ,
\end{equation}
which is valid only when,
\begin{equation*}
x{1} \neq \dfrac{9 f_{D} x_{3}-13 f_{D}+1}{2 f_{D}} \, .
\end{equation*}
The curve defined in Eq. (\ref{limcycle}) can be further specified
as follows:
\begin{itemize}
\item It becomes $x_{4}^{2} = \dfrac{-2 \left(\sqrt{-4 x_{1}+18
x_{3}-23}+5\right)-2 x_{1}+9 x_{3}}{6 x_{1}-27 x_{3}+36}$ in the
case of quasi-de Sitter expansion ($f_{D} = 1$) and when a
canonical scalar field is present ($c_{1} = -1$). The quantity on
the right-hand side is generally real and positive for $x_{1} >
\dfrac{1}{2} (9 x_{3}-16)$, so $x_{4}$ takes realistic values
across this curve. However, this curve is not proved to be an
invariant under the flow of the system, thus it can not be
categorized as an attractor. \item It becomes $x_{4}^{2} =
-\dfrac{\sqrt{15} \sqrt{-8 x_{1}+36 x_{3}-49}+6 x_{1}-27
x_{3}+33}{9 (6 x_{1}-27 x_{3}+38)}$ in the case of matter
domination ($f_{D} = 3$) and canonical scalar field ($c_{1} =
-1$). Again, the quantity on the right-hand side is generally
positive, so $x_{4}$ takes realistic values across the curve, for
$x_{1} > \dfrac{1}{2} (9 x_{3}-16)$. \item It becomes $x_{4}^{2} =
-\dfrac{5 \left(\sqrt{19} \sqrt{-88 x_{1}+396 x_{3}-533}+22
x_{1}-99 x_{3}+119\right)}{33 (22 x_{1}-99 x_{3}+138)}$ in the
case of radiation domination $\big( f_{D} = -\dfrac{11}{5} \big)$
and canonical scalar ($c_{1} = -1$). Once more, the quantity on
the right-hand side is generally positive, so $x_{4}$ takes
realistic values across the curve, for $x_{1} > \dfrac{1}{2} (9
x_{3}-16)$. \item It becomes $x_{4}^{2} = \dfrac{-\sqrt{16
x_{1}-72 x_{3}+89}+4 x_{1}-18 x_{3}+23}{6 x_{1}-27 x_{3}+33}$ in
the case of phantom scalar fields ($c_{1} = 1$ and $f_{D} =
\dfrac{1}{2}$). Here, the quantity on the right-hand side is
generally negative, so $x_{4}$ would take non-realistic complex
values across the curve. As a result, the attractor cannot exist
in the case of phantom scalar fields.
\end{itemize}
The above results indicate that canonical $k$-Essence $f(R)$
gravity has more appealing physical features quantified in the
presence of attractors in the phase space, for quite general
values of the free parameters.

\subsection{Possible Issues in the Model}

Before moving onto Model I$\beta$, we are bound to discuss a
couple of issues arising for Model I$\alpha$, that lead to
mathematically inconsistent or physically non-viable situations.

\subsubsection{Infinities for $c_{1}=1$}

Beginning from Eq. (\ref{fourthODEa}), we can clearly see the
existence of poles for $x_{4}^{2} = \pm \dfrac{c_{1}}{3 f_{D}}$
and eventually for,
\begin{equation}
x_{4} = \pm \sqrt{ \dfrac{c_{1}}{3 f_{D}} }  \, .
\end{equation}
Of course, these poles are purely imaginary for $c_{1} = -1$ and
are not so important in the case of canonical scalar fields.
However, assuming $c_{1} = 1$  and consequently $f_{D} >
\dfrac{1}{3}$, the poles are purely real and appear as two
straight hyperplanes along the phase space, in the form of,
\begin{equation*}
x_{4} = \pm \sqrt{ \dfrac{2}{3} } \, ,
\end{equation*}
assuming $f_{D} = \dfrac{1}{2}$ as we did in the visualization of
the vector field and the trajectories in the phase space. Close to
these hyperplanes, the derivative of $x_{4}$ tends to infinity,
and in effect the values of $x_{4}$ change rapidly towards
infinity (or minus infinity) as well. This behavior is not natural
and splits the phase space in three discrete and isolated
subspaces, each of which has a specific sink. Any initial values
of $x_{4} > \sqrt{ \dfrac{2}{3} }$ tends to $x_{4}^{*} = 1$ and
those with initial values for $x_{4} < -\sqrt{ \dfrac{2}{3} }$
tend to $x_{4}^{*} = -1$, while the initial conditions in $
-\sqrt{ \dfrac{2}{3} } < x_{4} < \sqrt{ \dfrac{2}{3} }$ tend to
$x_{4}^{*} = 0$. As a result, when a phantom scalar field is
present in contrast to the canonical scalar case, the state
variable $x_{4}$ may tend to zero through a stable manifold
\footnote{This was also demonstrated by the analytical solutions
of Eq. (\ref{fourthODEa})}. This indicates that $\dot{\phi} = 0$,
hence that the scalar field remains constant over time, is a
stable solution for our model. This once more indicates the
problematic physical situation that arises in the case where
phantom scalars are used. This was also demonstrated in Ref.
\cite{Nojiri:2019dqc} where the phantom scalar $k$-Essence $f(R)$
gravity theory had to be extremely fine tuned in order for it to
be viable and compatible with the Planck data.

\section{The Phase Space of the Model I$\beta$}

Now let us turn our focus on the model I$\beta$, and we shall
examine the phase space structure in the case $c_{1} = 0$. The
dynamical system describing such a cosmology is composed of Eqs.
(\ref{firstODE}, \ref{secondODE}, \ref{thirdODE}, \ref{fourthODEb}
and \ref{fifthODE}) and it is subjected to constraint
(\ref{FriedconstrainIb}) and the effective barotropic index of Eq.
(\ref{effeqstate}).

This system is similar to the previous, with only one differential
equation being altered namely Eq. (\ref{fourthODEb}), and some
terms being simplified, due to $c_{1} = 0$. As a result, many of
our previous statements (e.g. the integrability of Eq.
(\ref{thirdODE})) are still valid. However, some aspects of the
system are different, since this version is much simpler, for
example the number of critical points is reduced from sixteen to
four.

%%%%%%%%%%%%%%%%%%%%%%%%%%%%%%%%%%%%%%%%%%%%%%%%%%%%%%%%%%%%%%%%%%%%%%%%%%%%%%%%%%%%%%%%%%%%%%%%%%

\subsection{Stability of the Equilibrium Points}

Now let us investigate the stability of the fixed points by using
an alternative approach by utilizing divergence field
$\vec{V}(x_{1},x_{2},x_{3},x_{4},x_{5})$. Setting
$\vec{V}(x_{1},x_{2},x_{3},x_{4},x_{5}) =
0$,\footnote{$\vec{V}(x_{1},x_{2},x_{3},x_{4},x_{5})$ is the
vector field of the dynamical system, as defined in the previous
section.} we analytically derive four critical points, whose
coordinates are functions of $m$. Consequently, the equilibria of
the system are subjected to bifurcation, depending on the values
of parameters $m$. More specifically, the four critical points
have complex coordinates for $m>0$, so none of them is an
equilibrium point, and two of them arise with real coordinates for
$m \le 0$, so these two are equilibrium points. Both of these
equilibria fulfill both the Friedmann constraint (Eq.
(\ref{FriedconstrainIb}) and the effective equation of state, for
each specific $m$, thus both equilibria are viable cosmological
solutions. Since $m>0$ does not correspond to solutions with
physical meaning, we shall remain with $m \le 0$ and study the
three cases with realistic behavior, with $m=0$ for the quasi-de
Sitter expansion, $m=-\dfrac{9}{2}$ for the matter domination, and
$m=-8$ for the radiation domination.

The linearized system around a miscellaneous equilibrium point $\{
x_{i}^{*} \}$ is given as,
\begin{equation} \label{LinIb}
\small
\begin{pmatrix}
\dfrac{d \xi_{1}}{dN} \\ \dfrac{d \xi_{2}}{dN} \\ \dfrac{d
\xi_{3}}{dN} \\ \dfrac{d \xi_{4}}{dN} \\ \dfrac{d \xi_{5}}{dN}
\end{pmatrix}
 =
\begin{pmatrix}
 2 x_{1}^{*} - x_{3}^{*} + 3 & 0 & 2 - x_{1}^{*} & 2 f_{D} x_{4}^{3 \; *} x_{5}^{*} & \dfrac{f_{D} x_{4}^{4 \; *}}{2} \\
 x_{2}^{*} & x_{1}^{*} -2 x_{3}^{*} + 4 & -2 (x_{2}^{*} + 2) & 0 & 0 \\
 0 & 0 & 8 - 4 x_{3}^{*} & 0 & 0 \\
 0 & 0 & 0 & -1 & 0 \\
 -x_{5}^{*} & 0 & -2 x_{5}^{*} & 0 & -x_{1}^{*} - 2 x_{3}^{*} + 4 \\

\end{pmatrix}
\begin{pmatrix}
\xi_{1} \\ \xi_{2} \\ \xi_{3} \\ \xi_{4} \\ \xi_{5}
\end{pmatrix} \, ,
\end{equation}
\normalsize where $\xi_{i}$ the linear perturbations of the phase
space variables, $x_{i}$ around the equilibrium point.

Unlike the previous case, the $f_{D}$ parameter is not determined
from the value of $m$ via the constraint. Thus it can be taken as
a free parameter. Furthermore, the coordinates of the equilibrium
points do not depending on $f_{D}$, and it is removed from the
linearized system (since $x_{4}^{*} = x_{5}^{*} = 0$) and by
extent to the eigenvalues and eigenvectors of it close to the
equilibrium points. For simplicity, and without any loss of
generality, we set $f_{D} = 1$ for any numerical calculation or
plot we are about to conduct.

For a quasi-de Sitter evolution, $m = 0$, so we may derive the
following two equilibrium points,
\begin{equation}
P_{1} (-1,0,2,0,0) \;\;\; \text{and} \;\;\; P_{2}(0,-1,2,0,0) \, .
\end{equation}
Using the linearized system from Eq. (\ref{LinIb}), we may reach
to the following structure:
\begin{itemize}
\item[1] Point $P_{1}$ has three degenerate stable manifolds, in
the directions $v_{1} = (1,0,0,0,0)$, $v_{2} = (0,1,0,0,0)$ and
$v_{4} = (0,0,0,1,0)$, and one unstable manifold, in the direction
$v_{5} = (0,0,0,0,1)$; the remaining manifold, in direction
$v_{1,2,3} = (-3,4,-1,0,0)$ is central transitionary, since the
corresponding eigenvalue is zero. \item [2] Point $P_{2}$ has one
stable manifold, in direction $v_{4} = (0,0,0,1,0)$, and one
unstable manifold, in the direction $v_{1,2} = (-1,1,0,0,0)$; the
remaining three manifolds are central transitionary, since their
eigenvalues equal zero.
\end{itemize}
In the same way, for the matter dominated era, in which case
$m=-\dfrac{9}{2}$, the equilibrium points become,

Given $m = -\dfrac{9}{2}$, the equilibrium points become,
\begin{equation}
P_{1} \Big(
-\dfrac{5+\sqrt{73}}{4},\dfrac{7+\sqrt{73}}{4},\dfrac{1}{2},0,0
\Big) \;\;\; \text{and} \;\;\; P_{2} \Big(
-\dfrac{5-\sqrt{73}}{4},\dfrac{7-\sqrt{73}}{4},\dfrac{1}{2},0,0
\Big) \, .
\end{equation}
From the linearized system from Eq. (\ref{LinIb}), we obtain the
eigenvalues of each. Then the stability structure of the fixed
points are as follows,
\begin{itemize}
\item[1] Point $P_{1}$ has three stable manifolds, in the
directions $v_{1,2} = (-1,1,0,0,0)$, $v_{2} = (0,1,0,0,0)$ and
$v_{4} = (0,0,0,1,0)$, and two unstable manifolds. \item [2] Point
$P_{2}$ has two stable manifolds, in directions $v_{4} =
(0,0,0,1,0)$ and $v_{5} = (0,0,0,0,1)$, and three unstable
manifolds.
\end{itemize}
Finally, for the radiation domination era, given $m = -8$, the
equilibrium points are,
\begin{equation}
P_{1} (-4,5,0,0,0) \;\;\; \text{and} \;\;\; P_{2} (1,0,0,0,0) \, .
\end{equation}
From the linearized system from Eq. (\ref{LinIb}), the stability
of each is obtained as follows,
\begin{itemize}
\item[1] Point $P_{1}$ has two degenerate stable manifolds, in the
directions $v_{1,2} = (-1,1,0,0,0)$ and $v_{2} = (0,1,0,0,0)$, and
two degenerate unstable manifolds; the remaining one yields a zero
eigenvalue, being a central transitionary. \item [2] Point $P_{2}$
has one stable manifold, in directions $v_{4} = (0,0,0,1,0)$, and
four unstable manifolds; two of the latter have equal eigenvalues,
being degenerate.
\end{itemize}
Thus in the $c_1=0$ case, certainly the phase space contains some
stable attractors, and an interesting property of the phase space
is analyzed in the next section.

\subsection{A $1-d$ attractor}

\begin{figure}[h!] % "[t!]" placement specifier just for this example
\centering
\includegraphics[width=0.6\linewidth]{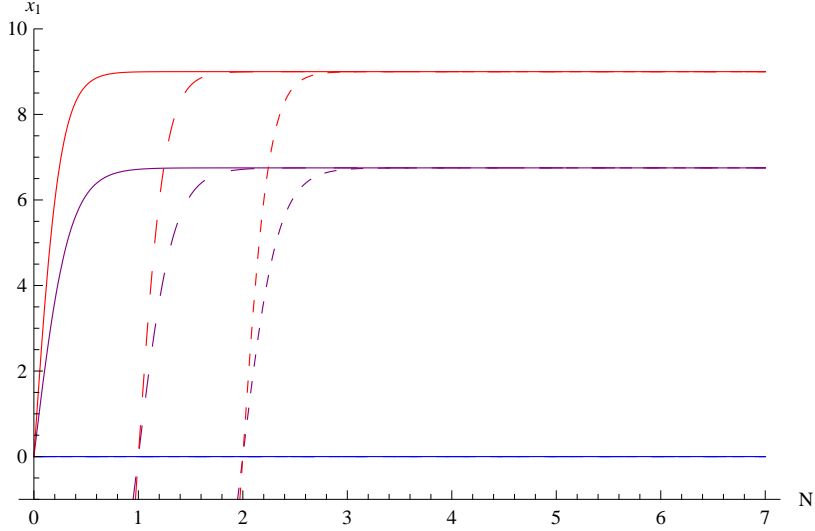}
\caption{Analytic solutions of Eq. (\ref{firstODE}) that
correspond to the $1-d$ attractor existing for $c_{1} = 0$. Blue
curves stand for $m=0$, purple curves for $m = -\dfrac{9}{2}$ and
red ones for $m = -8$; the solid curves denote initial conditions
for $N_{0} = 0$, while dashing becomes thinner as $N_{0} >0$
grows.} \label{fig:x1_attractor}
\end{figure}

As in the previous model, the flow of the system, defined as
$\vec{\nabla} \vec{V}$ does not maintain its sign, thus it is
impossible to define the system as conservative, dissipative or
explosive. More specifically,
\begin{equation} \label{flowIb}
\vec{\nabla} \vec{V} = 18 + 2 x_{1} - 9 x_{3} \, ,
\end{equation}
that changes sign astride a straight hypersurface. It is very
interesting that the flow of the system does not depend on the
variables $x_{2}$, $x_{4}$ and $x_{5}$ and on the parameters $m$
and $f_{D}$, thus the line is not subject to bifurcations as the
position and stability of the equilibrium points do. Demanding
that $\vec{\nabla} \vec{V} = 0$, we find that the equation of this
hypersurface (actually it is a line) is as follows,
\begin{equation} \label{attractor}
x_{1} = -\dfrac{9}{2}( 2 - x_{3} ) \, .
\end{equation}
This supersurface is fully determined as a function of the
$e$-foldings number, $N$, since $x_{3}$ is given as an analytic
solution of Eq. (\ref{thirdODE}). Thus, Eq. (\ref{attractor})
provides us with an analytic solution for $x_{1}$ as well, in the
form,
\begin{equation} \label{x1_analytical}
x_{1}(N) = -7 - \dfrac{ 9 \sqrt{-2m} }{4} \tan \big( \sqrt{-2m} (N
- N_{0}) \big) \, .
\end{equation}
Solutions of Eq. (\ref{firstODE}) are thus driven by solutions of
Eq. (\ref{thirdODE}), and by extent they are attracted towards the
value,
\begin{equation}
x_{1}^{\star} = -\dfrac{9 \sqrt{2m}}{4} \tan
\left(\sqrt{2m}\right) \, .
\end{equation}
This values does not correspond to an equilibrium point, except
for the quasi-de Sitter evolution case ($m=0$). Several analytic
solutions of Eq. (\ref{firstODE}) that correspond to the $1-d$
attractor existing for $c_{1} = 0$ are presented in Fig.
\ref{fig:x1_attractor}.

One should also give notice to the strong intermingling of the
$f(R)$ function and its rate of change, $\dot{f}(R)$, with the
curvature. If we substitute the variables $x_{1}$ and $x_{3}$ from
their definitions (Eq. (\ref{statevar}), to the Eq.
(\ref{attractor}), we obtain the following relation
\begin{equation}
\dfrac{\dot{f}}{f} = \dfrac{R}{6H} - 9H \, .
\end{equation}
Thus even the case $c_1=0$ provides us with a rich phase space
structure, although this case is less physically interesting in
comparison to the other two cases analyzed in the previous
sections.

%%%%%%%%%%%%%%%%%%%%%%%%%%%%%%%%%%%%%%%%%%%%%%%%%%%%%%%%%%%%%%%%%%%%%%%%%%%%%%%%%%%%%%%%%%%intro-conclusions-references-abstract-

\section{Conclusions and Discussion of the Results}

In this paper we thoroughly examined the phase space of a simple
$k$-Essence $f(R)$ gravity theory. We studied both the cases that
the $k$-Essence field consists of a phantom or a canonical scalar
field. We analyzed two models focusing on cosmological solutions
with physical interest and we emphasized our study on finding
physically interesting fixed points in the theory. After
appropriately choosing the phase space variables, we constructed
an autonomous dynamical system, with the only deviation from being
autonomous contained in the parameter $m=-\frac{\ddot{H}}{H^3}$.
It turned out that for cosmologically interesting cases, the
parameter $m$ takes constant values, and the dynamical system is
rendered autonomous. Specifically for $m=0$ it describes a
quasi-de Sitter cosmology, for $m=-\frac{9}{2}$ it describes a
matter dominated cosmology and finally for $m=-8$ it describes a
radiation domination era.

Proceeding to the analysis of the dynamical system, we isolated
three major cases, identified as $c_{1} = -1$ which corresponds to
a canonical scalar field, $c_{1} = 0$, which implies the absence
of a kinetic term for the scalar field, and $c_{1} = 1$ which
corresponds to phantom field cosmologies. These total nine
physical situations can be studied as different static versions of
a specific system. However, they can be studied as three different
systems (concerning the values of $c_{1}$) that are subject to a
bifurcation (concerning the values of $m$). In this sense, we may
extracted some interesting results, summarized below.

First of all, given that $c_{1} = -1$ or $c_{1} = 1$, the
bifurcation of $m$ from positive values to zero creates six
equilibrium points, and from zero to negative values to further
four; two of these four in the cases of canonical scalar fields,
and all four in the cases of phantom fields, are non-viable
equilibria, since they do not fulfill the Friedmann constraint
and/or the effective equation of state for the specific matter
fields content. Furthermore, the appearance of equilibrium points
occurs in such a way that specific symmetries are present and
easily observed in the values of $x_{1}$ and $x_{2}$ phase space
variables, even more in the values of $x_{4}$ and $x_{5}$ phase
space variables. Taking into account the definitions of the phase
space variables, the symmetries observed in $x_{1}$ and $x_{2}$
are symmetries concerning the specific form of the $f(R)$
function; on the other side, the symmetries observed in $x_{4}$
correspond to the behavior of the scalar field, $\phi$. Finally,
$x_{5}$ has a usual equilibrium value at zero, which denotes an
infinite rate of increase (or decrease) for either the Hubble
rate, or the $f(R)$ function. Intriguingly enough, the equilibrium
points for which $x_{5}^{*} \neq 0$ are usually the non-viable
points for $m < 0$. Furthermore, two of these are always found
asymptotically stable in four directions. The equilibria for which
$x_{5}^{*} = 0$ are usually found to be asymptotically unstable,
at least in directions such as $x_{5}$. This is the main source of
instability in the quasi-de Sitter case and quantifies
mathematically the ability of the present theory to generate the
graceful exit from inflation.

The four emerging (and partially non-viable) equilibrium points
are conceived as two pairs, mirrored on $x_{4} = 0$, which denotes
the constancy of the scalar field, since four of their coordinates
are exactly the same and the fifth (the $x_{4}^{*}$) takes
respectively the values $-1$ and $1$. In fact the stability of two
mirrors the stability of the other two, qualitatively (the number
of stable manifolds) and quantitatively (the direction of the
stable and unstable manifolds). The stability of each isolated
equilibrium is generally preserved through the bifurcations, with
one important note: moving from $m=0$ to $m<0$, the number of
central transitionary manifolds decreases and stable manifolds
replace them.

The remaining six (always viable) equilibrium points can also be
perceived as two groups of three, mirrored on $x_{4} = 0$, with
all other coordinates being equal (in each group), and $x_{4}$
taking the values $-1$, $0$ and $1$. The stability of each group
seems correlated, with points with relatively more stable
manifolds being grouped together and those with fewer stable
manifolds alike. In the same manner, the equilibria of a group
with $x_{4}^{*} = 0$ are proved to be relatively more unstable
that the other two. This peculiar symmetry reveals a fundamental
problem in the case of $\phi = \text{const.}$, that repels
solutions towards $x_{4}^{*} = -1$ or $x_{4}^{*} = 1$. The gradual
disappearance of central transitionary manifolds as $m$ moves from
zero to negative values is observed here as well. Following the
typical scheme for the evolution of the Universe, some of the
central manifolds are preserved when $m = -8$ (radiation-dominated
era) but are completely absent for $m = -\dfrac{9}{2}$, proving
that the values of parameter $m$ are not decreasing linearly and
uniformly.

Given the degenerate case of $c_{1} = 0$, the number of
equilibrium points is narrowed down to only two. Here $x_{4}^{*} =
x_{5}^{*} = 0$ always. Still, both equilibria are proved viable
according to the Friedmann constraint and the effective equation
of state. The two equilibrium points generally preserve their
stable manifolds as the values of $m$ move from $0$ to
$-\dfrac{9}{2}$ and $-8$. The first of these points for which
$x_{1}^{*} < x_{2}^{*}$, has more stable manifolds. Once more, the
transition from $m = 0$ to $m = -8$ does not completely transform
the central transitionary manifolds to stable ones, as it happens
when $m = -8$ changes to $m = -\dfrac{9}{2}$. The standard model
for cosmic evolution is somehow present in this phase transition,
concerning the stability of the equilibria.

Second, a possible attractor appears in all three major cases. For
$c_{1} = 0$ the attractor is a 1-d hypersurface, that connects the
$x_{1}$ and $x_{3}$ phase space variables and does not depend on
any parameter values. When $c_{1} = -1$ or $c_{1} = 1$, the
attractor is a 1-d hypersurface, that connects the $x_{1}$,
$x_{3}$ and $x_{4}$ phase space variables and depends strongly on
the choice of $f_{D}$ and $c_{1}$. However, its complexity is such
that its existence is not guaranteed. Interestingly, in any of the
above cases the possible attractor connects the rate of change of
the $f(R)$ function (from the $x_{1}$ variable) to the curvature
scalar $R$ (from the $x_{3}$ variable) and to the rate of change
of the scalar field (from the $x_{4}$ variable). However, the two
of these variables do not only correspond to specific symmetries
in the model, as observed in the equilibrium points, but also
determine the effective equation of state and the Friedmann
constraint respectively. Furthermore, their dynamical equations
are separated from the other variables in a certain way, leading
to their integrability and triviality.

Finally, if we observe this separation of Eqs. (\ref{thirdODE})
and (\ref{fourthODEa} or \ref{fourthODEb}) from the other three
and analyze their integrability, we come to the easy result that
both $x_{3}$ and $x_{4}$, or with other words the scalar curvature
and the rate of change of the scalar field, are trivially and
independently evolving towards an equilibrium value. As long as
the scalar field is concerned, the attained equilibrium value(s)
are normal and correspond to viable cosmological solutions. When
the scalar curvature is taken into account, we observe that aside
from the case of quasi-de Sitter evolution, the equilibrium
attained does not correspond to the barotropic index of the
specific matter fields content, and thus relates to non-viable
cosmological solutions.

As a consequence of the above, the system is characterized by two
extremely different states. On the one hand, an asymptotic
instability is observed in almost every equilibrium point that was
noted. This instability is further amplified, if we take into
consideration that the greater stability resolves around
non-viable cosmological solutions. Concerning quasi-de Sitter
fixed points, this asymptotical instability after an attractor is
reached, may be viewed as an inherent mechanism for graceful exit
in the $k$-Essence $f(R)$ gravity theory.

On the other hand, a high degeneracy is easy to be noted, given
the fact that many equilibrium points emerge, and have specific
symmetries in their coordinates, as well as the eigenvalues and
eigenvectors of the linear perturbations, that characterize their
stability. Also two of the dynamical equations are separated from
the rest and integrated, resulting to trivial solutions for
$x_{3}$ and $x_{4}$. The third state variable, $x_{1}$, is (in
some cases) strongly interconnected with the $x_{3}$ and $x_{4}$
via an attractor, and thus its behavior is analytically traced and
found diverging from the noted equilibria (viable or not).
Finally, many of the stable or unstable manifolds around the
equilibrium points are proved degenerate.

A possible resolution of this degeneracy would be the
transformation of the system into another and the subsequent
reduction of its dimensions from five to four, or ever three. This
would expel the degeneracy and would give us a clearer picture for
the general behavior of the system, under the aforementioned
bifurcations. However, the fundamental instabilities of the system
are not expected to alter following such a transformation.  This
issue is more probably an issue of the theoretical framework out
of which the models were derived, rather than of the specific
dynamical system, indicating for the quasi-de Sitter fixed points
that the final attractors are unstable asymptotically and thus
this is a strong hint that the theory possesses an internal
structure that allows a graceful exit from inflation.

\end{document}